\documentclass{emulateapj}

\usepackage{amssymb}
\usepackage{times}
\usepackage{amsmath}
\usepackage{subfigure}
\usepackage{multirow}
\usepackage[colorlinks,backref,dvipdfm,linkcolor=blue,anchorcolor=,citecolor=blue]{hyperref}
\makeatletter

\newcommand{\rmnum}[1]{\romannumeral #1}
\newcommand{\Rmnum}[1]{\expandafter\@slowromancap\romannumeral #1@}
\makeatother

\slugcomment{}

\shorttitle{\textit{High-precision asteroseismology of HD 50230}}
\shortauthors{\textit{Wu \& Li}}

\begin{document}


\title{High-precision asteroseismology in Slowly Pulsating B star: HD 50230}


\author{
Tao Wu\altaffilmark{1,2,3} and Yan Li \altaffilmark{1,2,3,4}
}


\altaffiltext{1}{Yunnan Observatories, Chinese Academy of Sciences, 396 Yangfangwang, Guandu District, Kunming, 650216, P. R. China; wutao@ynao.ac.cn, ly@ynao.ac.cn}
\altaffiltext{2}{Key Laboratory for the Structure and Evolution of Celestial Objects, Chinese Academy of Sciences, 396 Yangfangwang, Guandu District, Kunming, 650216, P. R. China}
\altaffiltext{3}{Center for Astronomical Mega-Science, Chinese Academy of Sciences, 20A Datun Road, Chaoyang District, Beijing, 100012, P. R. China}
\altaffiltext{4}{University of Chinese Academy of Sciences, Beijing 100049, China}


\begin{abstract}
  SPB star HD 50230, in fact a hybrid B-type pulsator, has been observed by CoRoT least 137 days. A nearly equidistant period spacing pattern are found among 8 modes which are extracted from the oscillation spectrum with more than 500 frequencies. However, it is thought to be most likely accidental by \citet[][]{Szewczuk2014IAUS}.
  In the present work, we analyze the 8 modes in depth with the $\chi^2$-matching method.
  Based on the best fitting model (model MA), we find that they can be well explained as a sequences of consecutive dipolar $(l,~m)=(1,~0)$. The period discrepancies between observations and the best fitting model are within 100 s except for the outlier which is up to 300 s. Based on the calculated CMMs, we find that, for pure g-mode oscillations, the buoyancy radius $\Lambda_{0}$ can be precisely measured with  the $\chi^2$-matching method between observations and calculations. It represents the ``Propagation time" of the g-mode from stellar surface to center. It is of $\Lambda_{0}=245.78\pm0.59~\mu$Hz with a precision of 0.24\%.
  In addition, we also find that HD 50230 is a metal-rich ($Z_{\rm init}=0.034-0.043$) star with a mass of $M=6.15-6.27~{\rm M_{\odot}}$. It is still located on hydrogen-burning phase with central hydrogen $X_{\rm C}=0.298-0.316$ (or $X_{\rm C}=0.306^{+0.010}_{-0.008}$), therefore has a convective core with a radius of $R_{\rm cc}=0.525-0.536~{\rm R_{\odot}}$ (or $R_{\rm cc}=0.531^{+0.005}_{-0.006}~{\rm R_{\odot}}$). In order to well interpret the structure of observed period spacing pattern, the convective core overshooting ($f_{\rm ov}=0.0175-0.0200$) and the extra diffusion mixing ($\log D_{\rm mix}=3.7-3.9$) should be taken into account in theoretical models.
\end{abstract}



\section{Introduction}\label{sec-intr}

Slowly pulsating B stars (hereafter SPB stars) are the upper main-sequence stars of intermediate mass ($2.5\sim8~M_{\odot}$) \citep[more descriptions see e.g.,][]{Aerts2010}. Their effective temperatures range from about 11000 to 22000 K with spectral type between B3 and B9. Their pulsation pattern represents the characteristic of non-radial, high-order, low-degree multi-periodic g-mode oscillations with the period of about $0.5-3$ days \citep{Aerts2010}. The oscillations of SPB stars are thought as excited by the $\kappa$-mechanism, i.e., the opacity is enhanced due to the ionization of iron-group elements, also called the $Z$-bump, at the temperature of about 200,000 K ($\log T\sim5.3$) \citep[see e.g.,][]{Aerts2010,Moravveji2016ApJ}.

Recently, more and more SPB stars have been observed and abundant seismological data have been collected {\bf \citep[e.g.,][]{Degroote2010Natur,Degroote2012AA, Papics2012AA,Papics2014AA,Papics2015ApJL,Papics2017AA, Moravveji2015AA,Moravveji2016ApJ,Triana2015ApJ, Briquet2016A&A, Zhang2018ApJ, Bowman2018arXiv, Buysschaert2018arXiv,Pedersen2019ApJ} via ground- and space-based missions, such as {\it CoRoT} \citep[e.g.,][]{Baglin2006cosp}, {\it Kepler} \citep[e.g.,][]{Borucki2010Sci,Koch2010ApJ,Gilliland2010PASP}, {\it K2} \citep[e.g.,][]{Haas2014AAS,Howell2014PASP}, and {\it TESS} \citep[e.g.,][]{Ricker2015JATIS}.} Some of SPB stars have extremely slow rotation, such as KIC 10526294 with a rotational period of $P_{\rm rot}\sim188$ days \citep[e.g.,][]{Papics2014AA,Moravveji2015AA}; some of them are moderate or faster rotators, such as KIC 7760680 with a rotational frequency of $f_{\rm rot}\sim0.48~{\rm day}^{-1}$ \citep[][]{Moravveji2016ApJ}, KIC 3459297, KIC 6352430A, KIC 4930889 (A?), KIC 9020774, and KIC 11971405 with the rotational frequencies of $f_{\rm rot}\sim0.63$, $0.64$, $0.74$, $1.06$, and $1.62~{\rm day}^{-1}$, respectively \citep[see][refer to Table 14]{Papics2017AA}. In addition, some SPB stars have both of rotation and magnetic field, such as $\zeta$ Cassiopeiae has a moderate rotation period of $P_{\rm rot}=5.370447\pm0.000078$ days and a weakly magnetic field whose intensity is about $100-150 ~{\rm G}$ \citep[see][]{Briquet2016A&A}; HD 43317 has a lower rotation period of $P_{\rm rot}=0.897673$ days and a magnetic field of $B_{p}=1312\pm332 ~{\rm G}$ \citep[see][]{Buysschaert2018arXiv}.

Recently, \citet[][]{Moravveji2015AA} made detailed model analyses for the ultra-slow rotating SPB star KIC 10526294. Finally, they found that the profile of elements, i.e., the shape of buoyancy frequency $N$, beyond and near the convective core can be constrained by comparing theoretical models with the observed 19 dipole prograde $(l,~m)=(1,~0)$ g-modes through period spacing pattern (i.e., period spacing vs. period). In the process, they also found that the exponentially decaying diffusion in convective core overshooting ($f_{\rm ov}=0.017-0.018$) is better than the corresponding step function formulation, in addition, an extra diffusion mixing ($\log D_{\rm mix}=1.75-2.00$) is necessary for explaining the sine-like structure in the observed period spacing pattern in this target. 

\citet[][]{Moravveji2016ApJ} analyzed 37 clearly identified dipole prograde $(l,~m)=(1,~+1)$ g-modes in KIC 7760680. Similar to \citet[][]{Moravveji2015AA} in KIC 10526294, they as well found that the exponentially decaying diffusion in convective core overshooting is better than the corresponding step function formulation. In addition, they found that the convective core overshooting ($f_{\rm ov}\thickapprox0.024\pm0.001$) can coexist with moderate rotation ($f_{\rm rot}\sim0.48~{\rm day}^{-1}$). In order to interpret the sine-like structure of period spacing pattern, the extra diffusion mixing is also necessary in theoretical models. But, they found the optimal extra diffusion mixing coefficient of $\log D_{\rm mix}\thickapprox0.75\pm0.25$ is notably smaller than that of KIC 10526294 ($\log D_{\rm mix}=1.75-2.00$), in that KIC 7760680 
rotates faster 
and also has older age compared to KIC 10526294.

\citet[][]{Buysschaert2018arXiv} modelled HD 43317 via 16 identified modes which are retrograde modes with $(l,~m)=(1,~-1)$ and $(2,~-1)$ and one distinct prograde $(2,~2)$ mode and constrained the surface strength of magnetic filed ($B_{p}=1312\pm332~{\rm G}$), except for precisely determining its fundamental parameters. On the other hands, they found that the magnetic field cause a suppression of near-core mixing in this star and then it has a lower convective core overshooting parameter ($f_{\rm ov}=0.004^{+0.014}_{-0.002}$).

These previous works, \citet[][]{Moravveji2015AA}, \citet[][]{Moravveji2016ApJ}, and \citet[][]{Buysschaert2018arXiv}, give us so much new insights in seismically modelling SPB stars and make us on a higher level to learn the interior structure of SPB stars, such as the connection among the extra diffusive mixing, the shape of buoyancy frequency $N$, and the structure of period spacing pattern, the relationship and/or interaction among convective core overshooting, rotation, and magnetic field.

However, there are still significant discrepancies between observed periods and model calculated ones. The maximum period discrepancy is up to about 500 s for both of KIC 10526294 \citep[see][Figure 4]{Moravveji2015AA} and KIC 7760680 \citep[see][Figure 8]{Moravveji2016ApJ}. For HD 43317, the period discrepancy between observations and best fitting model is up to thousands seconds \citep[see][Table 3]{Buysschaert2018arXiv}. So, what factors is such large period discrepancy between the observations and the best fitting models caused by? Up to now, the question is still hung in the sky.

In the present work, we will make detail seismic analyses for an unique SPB star HD 50230 to probe its interior structures. HD 50230 has a larger mass ($7-8~{\rm M_{\odot}}$) \citep[][]{Degroote2010Natur}, which is close to the upper limit of the mass range of SPB stars ($2.5-8~{\rm M_{\odot}}$) and larger than that of the other analyzed SPB stars \citep[refer to e.g.,][]{Moravveji2015AA,Moravveji2016ApJ,Buysschaert2018arXiv}. It also has higher metallicity ($\log Z/Z_{\odot}=0.3$) and stays in a binary system (more detail introduction about HD 50230 seeing the next section).

\section{HD50230}\label{sec.HD5}
\begin{deluxetable*}{lcccc}
\tablecaption{Summary of observations for HD 50230.\label{table_obs}}
\tablehead{
\colhead{Fre. ID}& \colhead{$n_{g}$}& \colhead{$\nu_{i}\pm\sigma_{\nu_{i}}^{\rm a}$} & \colhead{$P_{i}\pm\sigma_{P_{i}}^{\rm a}$}  & \colhead{$\Delta P_{i}\pm\sigma_{\Delta P_{i}}^{\rm a}$}  \\
\colhead{}         & \colhead{}         & \colhead{[$10^{-3}$day$^{-1}$]}     & \colhead{[s]}      & \colhead{[s]}
}
\startdata
$f_{028}$ & $-6$ & $1236.85\pm0.04$ & $69854.873\pm6.777$   & $9241.379\pm20.692$ \\
$f_{101}$ & $-7$ & $1092.34\pm0.09$ & $79096.252\pm19.551$  & $9640.178\pm20.301$ \\
$f_{006}$ & $-8$ & $973.67\pm0.02$ & $88736.430\pm5.468$   & $9522.432\pm8.652$ \\
$f_{005}$ & $-9$ & $879.31\pm0.02$ & $98258.862\pm6.705$   & $9186.452\pm10.451$ \\
$f_{011}$ & $-10$ & $804.13\pm0.02$ & $107445.314\pm8.017$  & $9562.853\pm12.436$ \\
$f_{016}$ & $-11$ & $738.41\pm0.02$ & $117008.166\pm9.508$ & $9233.798\pm14.590$ \\
$f_{001}$ & $-12$ & $684.40\pm0.02$ & $126241.964\pm11.067$ & $9430.037\pm33.819$ \\
$f_{052}$ & $-13$ & $636.83\pm0.05$ & $135672.000\pm31.956$ & \nodata
\enddata
\tablenotetext{}{Notes: Observations include frequencies $\nu_{i}$, and associated periods ($P_{i}$) and period spacings ($\Delta P_{i}$) and the corresponding observational uncertainties $\sigma_{\nu_{i}}$, $\sigma_{P_{i}}$, and $\sigma_{\Delta P_{i}}$, respectively. The radial order ($n_{g}$) is decided from the best fitting model.}
\tablenotetext{a}{These observational frequencies and the corresponding observational uncertainties ($\nu_{i}\pm\sigma_{\nu_{i}}$) come from \citet[][Table A.2]{Degroote2012AA}. Periods ($P_{i}$) and period spacings ($\Delta P_{i}$) are derived from the frequencies ($\nu_{i}$). Similar to Table 1 of \citet[][]{Degroote2012AA}, the uncertainties of periods and period spacings ($\sigma_{P_{i}}$ and $\sigma_{\Delta P_{i}}$) correspond to $3\sigma_{\nu_{i}}$ in the table.}
\end{deluxetable*}

HD 50230 is a metal-rich ($\log Z/Z_{\odot}=0.3$ dex; $Z_{\odot}=0.02$) young massive star with a spectral type of B3V and a visual magnitude of 8.95 \citep[e.g.][]{Degroote2010Natur,Degroote2012AA,Szewczuk2014IAUS}. It is the primary component of a binary system with the rotational velocity ($V_{\rm eq}\sin i$)  of $6.9\pm1.5$ km s$^{-1}$. The effective temperature is of $T_{\rm eff}=18000\pm1500$ K, and surface gravity to be of $\log g=3.8\pm0.3$ (c.g.s. units) \citep[refer to][]{Degroote2012AA}. For the effective temperature of the secondary component, \citet[][]{Degroote2012AA} given an upper limit of $T_{\rm eff,2}\leqslant16000$ K when assuming surface gravity $\log g_{2}\thickapprox4$ (c.g.s. units).

Based on CoRoT observations of $\sim137$ days, \citet[][]{Degroote2010Natur} extracted eight peaks, which is listed in Table \ref{table_obs}, from the oscillation power spectra. These peaks are almost uniformly spaced in period with the mean spacing of 9418 s and a deviation of about 200 s. \citet[][]{Degroote2010Natur} interpreted them as consecutive radial orders, $n$, with the same spherical harmonic degree, $l$, and azimuthal order, $m$. Based on the assumption of $l=1$ and $m=0$, \citet[][]{Degroote2010Natur} found the mass of HD 50230 to be of $7-8$ ${\rm M_{\odot}}$ with an overshooting extension of $\alpha_{\rm ov}\geqslant0.2H_{P}$ ($H_{P}$ pressure
scale height) by the step overshooting description in the radiative regions adjacent to the convective core \citep[][for an overview]{Szewczuk2014IAUS}. In addition, \citet[][]{Degroote2010Natur} suggested the HD 50230 is staying in the middle of the main sequence. And about 60\% initial hydrogen in its center has already been consumed.

Motivated by the detection of period spacings, \citet[][]{Degroote2012AA} reanalyzed the observations. They extracted 566 frequencies from the oscillation power spectra, including high- and low-order g-mode and low-order p-mode. Most of them are not clearly identified. Based on those extracted frequencies, they obtained a rotational splitting of p-mode to be of $\Delta f_{\rm obs}=0.044\pm0.007 ~{\rm day}^{-1}$ via the analysis of the autocorrelation of the periodogram between 10 and 15 day$^{-1}$. Finally, \citet[][]{Degroote2012AA} suggested the rotational effects should be take into account in modelling when interpreting the small deviations from the uniform period spacings in despite of the fact that the surface rotational velocity is merely the order of magnitude of 10 km s$^{-1}$.

Recently, \citet[][]{Szewczuk2014IAUS} reanalyzed the light curves of HD 50230 and extracted 515 frequencies to try to interpret the oscillation spectra in-depth and to re-identify the detected frequencies. They found three series modes nearly uniformly spaced in periods from these extracted frequencies. But, they have some common modes. For the largest series, it has 11 frequencies and includes 8 frequencies of \citet[][]{Degroote2010Natur,Degroote2012AA} determined \citep[see Figure 1 of][]{Szewczuk2014IAUS}. But, finally, the 11 frequencies are thought as different degrees $l$ and azimuthal orders $m$.

According the equatorial velocity to be the order of magnitude of $V_{\rm eq}\sim10$ km s$^{-1}$ \citep[][]{Szewczuk2014IAUS}, assuming the radius be of $R\sim5~{\rm R_{\odot}}$, the corresponding rotational frequency is about $\Omega_{\rm rot}\equiv V_{\rm eq}/R\sim0.040$ day$^{-1}$ which is far smaller than the oscillation frequencies $\nu\sim1$ day$^{-1}$ (see Table \ref{table_obs}), i.e., $\Omega_{\rm rot}\ll\nu$. It indicates that the perturbation theory of rotational splitting for low-degree, high-order modes is available for the slow rotating star HD 50230.

Based on the rotational splitting of p-mode $\Delta f_{\rm obs}=0.044\pm0.007 ~{\rm day}^{-1}$, and the first-order approximation of rotational splitting of low-degree, high-order modes \citep{Dziembowski1992ApJ,jcd2003,Aerts2010}:
\begin{equation}\label{eq_domega2}
\begin{split}
&\delta\omega_{{\rm rot},nlm}^{\rm I} \approx m\Omega_{\rm rot,s}\left(1-\frac{1}{L}\right), ~~~~{\rm for ~g~mode};\\
&\delta\omega_{{\rm rot},nlm}^{\rm I} \approx m\Omega_{\rm rot,s}, ~~~~{\rm for ~p~mode},
\end{split}
\end{equation}
where $L=l(l+1)$, the corresponding rotational splitting of g-mode is about $\Delta f_{{\rm obs,g,}m=1}^{\rm I}\approx0.022\pm0.0035 ~{\rm day}^{-1}$ for dipolar modes. Correspondingly, the second-order approximation term of the low-degree, high-order mode is about $\Delta f_{{\rm obs,g,}m=1}^{\rm II}\approx1.2\times10^{-5} ~{\rm day}^{-1}$ (the second-order approximation term $\delta\omega_{{\rm rot},nlm}^{\rm II} \approx -\frac{m^2\Omega_{\rm rot,s}^2}{\omega_{nl0}}\frac{4L(2L-3)-9}{2L^2(4L-3)}$ is derived from \citet[][]{Dziembowski1992ApJ}). The second-order approximation term can be ignored compared to the first-order approximation term $\Delta f_{{\rm obs,g,}m=1}^{\rm I}$ for target HD 50230. In addition, combining the rotational splitting of p-mode and Equation \eqref{eq_domega2}, we obtain the rotational frequency of HD 50230 to be of $\Omega_{\rm rot,s}\backsimeq0.044\pm0.007~{\rm day}^{-1}$.

\section{Physical inputs and modelling}

\subsection{Physical inputs}
In the present work, our theoretical models were computed by the Modules of Experiments in Stellar Astrophysics (MESA), which was developed by \citet{MESA2011}. It can be used to calculate both the stellar evolutionary models and their corresponding oscillation information \citep{MESA2013,MESA2015}. We adopt the package {\small \textbf{``pulse"}} of version \small ``v6208" to make our calculations for both stellar evolutions and oscillations \citep[for more detailed descriptions refer to][]{MESA2011,MESA2013,MESA2015,MESA2016,MESA2018}. The package {\small \textbf{``pulse"}} is a test suite example of MESA in the directory of `\$MESA\_DIR/star/test\_suite/pulse'. The module of pulsation calculation is based on the ADIPLS code, which was developed by \citet[][]{jcd2008} and added into MESA by MESA team \citep[more information refer to][]{MESA2011,MESA2013,MESA2015,MESA2016,MESA2018}.

Based on the default parameters, we adopt the OPAL opacity table GS98 \citep{gs98} series. We choose the Eddington grey-atmosphere $T-\tau$ relation as the stellar atmosphere model, and treat the convection zone by the standard mixing-length theory (MLT) of \citet{cox1968} with mixing-length parameter $\alpha_{\rm MLT}=2.0$.

In the previous asteroseismic modelling of SPB stars, \citet[][]{Moravveji2015AA} and \citet[][]{Moravveji2016ApJ} reported that the exponentially decaying diffusive description is better than a step function formulation for treating the convective core overshooting. Here, we also adopt the theory of \citet[][]{Herwig2000} to treat the convective overshooting in the core. The overshooting mixing diffusion coefficient $D_{\rm ov}$ exponentially decreases with distance which extends from the outer boundary of the convective core with the Schwarzschild criterion:
\begin{equation}\label{eq_Dov}
D_{\rm ov}=D_{\rm conv,0}\exp{\left(-\frac{2z}{f_{\rm ov}H_{P,0}}\right)},
\end{equation}
where $D_{\rm conv,0}$ and $H_{P,0}$ are the MLT derived diffusion coefficient near the Schwarzschild boundary and the corresponding pressure scale height at that location, respectively. $z$ is the distance in the radiative layer away from that location. $f_{\rm ov}$ is an adjustable parameter \citep[for more detailed discriptions refer to][]{Herwig2000,MESA2011}.

In addition, the element diffusion, semi-convection, thermohaline mixing, and the mass-loss were not included in the theoretical models.

The previous works, \citet[][]{Degroote2010Natur}, \citet[][]{Moravveji2015AA}, and \citet[][]{Moravveji2016ApJ}, suggested that the extra diffusion mixing ($D_{\rm mix}$) is necessary in theoretical models for interpreting the deviations of period spacings.
In the present work, we also take it into account in the theoretical models.

Similar to the works of \citet[][]{Moravveji2015AA} and \citet[][]{Moravveji2016ApJ}, we also set the initial hydrogen mass fraction $X_{\rm init}=0.71$ which takes from the Galactic B-star standard \citep[][]{Nieva2012AA}. Therefore, the adjustable parameter of element compositions reduce to one, initial metal mass fraction $Z_{\rm init}$ or helium mass fraction $Y_{\rm init}$. They follow the relation: $X_{\rm init}+Y_{\rm init}+Z_{\rm init}=1$.

\subsection{Modelling and Finding the Best Fitting Model}

According to the above described, there are four initial input parameters, stellar mass ($M$), initial metal mass fraction ($Z_{\rm init}$), overshooting parameter in convective core ($f_{\rm ov}$) and the extra diffusion mixing ($\log D_{\rm mix}$) in theoretical models. The ranges and corresponding steps of the initial input parameters are listed in Table \ref{table_mg}. According to the study of \citet[][]{Degroote2010Natur}, \citet{Degroote2012AA}, and \citet{Szewczuk2014IAUS}, we preliminarily set stellar mass $M\in[6.6,~8.2]~{\rm M_{\odot}}$ with a step of 0.2 ${\rm M_{\odot}}$, initial metal mass fraction $Z_{\rm init}\in[0.010,~0.040]$ with a step of 0.005, overshooting parameter $f_{\rm ov}\in[0.010,~0.035]$ with a step of 0.005, and the extra diffusion mixing coefficient $\log D_{\rm mix}\in[2.0,~4.5]$ with a step of 0.5, respectively (Grid A hereafter).

\begin{deluxetable}{lcc}
\tablecaption{Model calculation grids. \label{table_mg} }
\tablehead{
\colhead{Variables} & \colhead{Parameter Ranges} & \colhead{Steps}
}
\startdata
\multicolumn{3}{c}{Grid A}\\
\hline

\hline
Mass ($M/{\rm M_{\odot}}$) & $6.6-8.2$  & 0.2 \\
Initial metal abundance ($Z_{\rm init}$) & $0.010-0.040$  & 0.005 \\
Overshooting parameter ($f_{\rm ov}$) & $0.010-0.035$     & 0.005 \\
Extra mixing ($\log D_{\rm mix}$) & $2.0-4.5$   & 0.5 \\
\hline

\hline
\multicolumn{3}{c}{Grid B}\\
\hline

\hline
Mass ($M/{\rm M_{\odot}}$) & $6.0-7.2$  & 0.1 \\
& $6.10-6.30$  & 0.05 \\
Initial metal abundance ($Z_{\rm init}$) & $0.0300-0.0425$  & 0.0025 \\
Overshooting parameter ($f_{\rm ov}$) & $0.0150-0.0250$     & 0.0025 \\
Extra mixing ($\log D_{\rm mix}$) & $3.2-4.4$   & 0.2 \\
& $3.4-4.2$  & 0.1 \\
\hline

\hline
\multicolumn{3}{c}{Grid C}\\
\hline

\hline
Mass ($M/{\rm M_{\odot}}$) & $6.1500-6.3000$  & 0.0125 \\
Initial metal abundance ($Z_{\rm init}$) & $0.03500-0.04375$  & 0.00125 \\
Overshooting parameter ($f_{\rm ov}$) & $0.0150-0.0225$     & 0.0025 \\
 &  $0.01750-0.02000$  & 0.00125 \\
Extra mixing ($\log D_{\rm mix}$) & $3.6-4.2$   & 0.1
\enddata
\end{deluxetable}

\begin{figure}
  \begin{center}
  \includegraphics[scale=0.33,angle=-90]{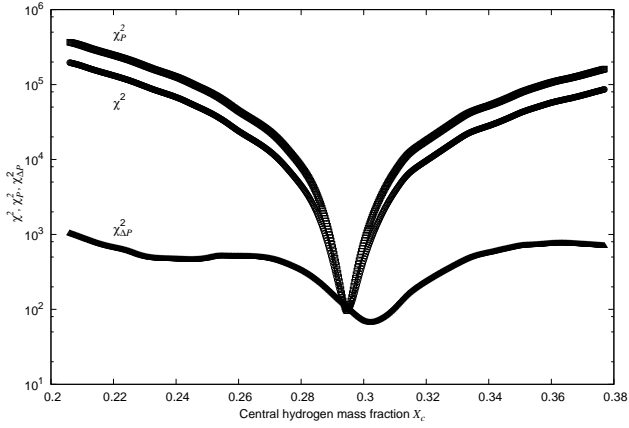}
   \caption{$\chi^2$, $\chi^2_{P}$, and $\chi^2_{\Delta P}$ as a function of central hydrogen mass fraction (central hydrogen hereafter) $X_{\rm C}$ for the evolution of stellar mass $M=6.20 ~{\rm M_{\odot}}$, initial metal mass fraction $Z_{\rm init}=0.040$, initial hydrogen mass fraction $X_{\rm init}=0.71$, overshooting parameter $f_{\rm ov}=0.020$, and extra mixing coefficient $\log D_{\rm mix}=3.8$.
  }\label{fig.chi}
  \end{center}
\end{figure}

Based on the above initial input parameters, we use MESA to calculate the corresponding stellar models and oscillation frequencies. Similar to the work of \citet[][]{wu2016ApJL}, \citet[][]{wu2017ApJ}, and \citet[][]{wu2017EPJWC}, we will use $\chi^2$-matching method to compare the observations and models to search the best fitting model. In the process, we also merely use the asteroseismic information (periods and/or period spacings) to constrain the theoretical models. Firstly, we decide the $\chi^2$-minimization model (CMM hereafter) from every evolutionary tracks as shown in Figure \ref{fig.chi}. Finally, decide the best fitting model from the selected CMMs. The $\chi^2$ is defined as:
\begin{equation}\label{eq:chiP}
\begin{split}
\chi^2_{P}=&\frac{1}{N}\sum^{N}_{i=1}\left( \frac{P^{\rm obs}_{i}-P^{\rm mod}_{i}}{\sigma_{P^{\rm obs}_{i}}}\right)^2,\\
\chi^2_{\Delta P}=&\frac{1}{N-1}\sum^{N-1}_{i=1}\left( \frac{\Delta P^{\rm obs}_{i}-\Delta P^{\rm mod}_{i}}{\sigma_{\Delta P^{\rm obs}_{i}}}\right)^2,\\
\chi^2=&\frac{1}{2N-1}\left[N\chi^2_{P}+(N-1)\chi^2_{\Delta P} \right],
\end{split}
\end{equation}
where $N=8$, the superscripts ``obs" and ``mod" represent observations and theoretical calculations, respectively. $P_{i}$ and $\Delta P_{i}$ ($=P_{i+1}-P_{i}$) are the oscillation period and period spacing, respectively. $\sigma_{P^{\rm obs}_{i}}$ and $\sigma_{\Delta P^{\rm obs}_{i}}$ denote their corresponding observational errors, which are listed in Table \ref{table_obs}.

\begin{figure*}
  \begin{center}
  \includegraphics[scale=0.3,angle=-90]{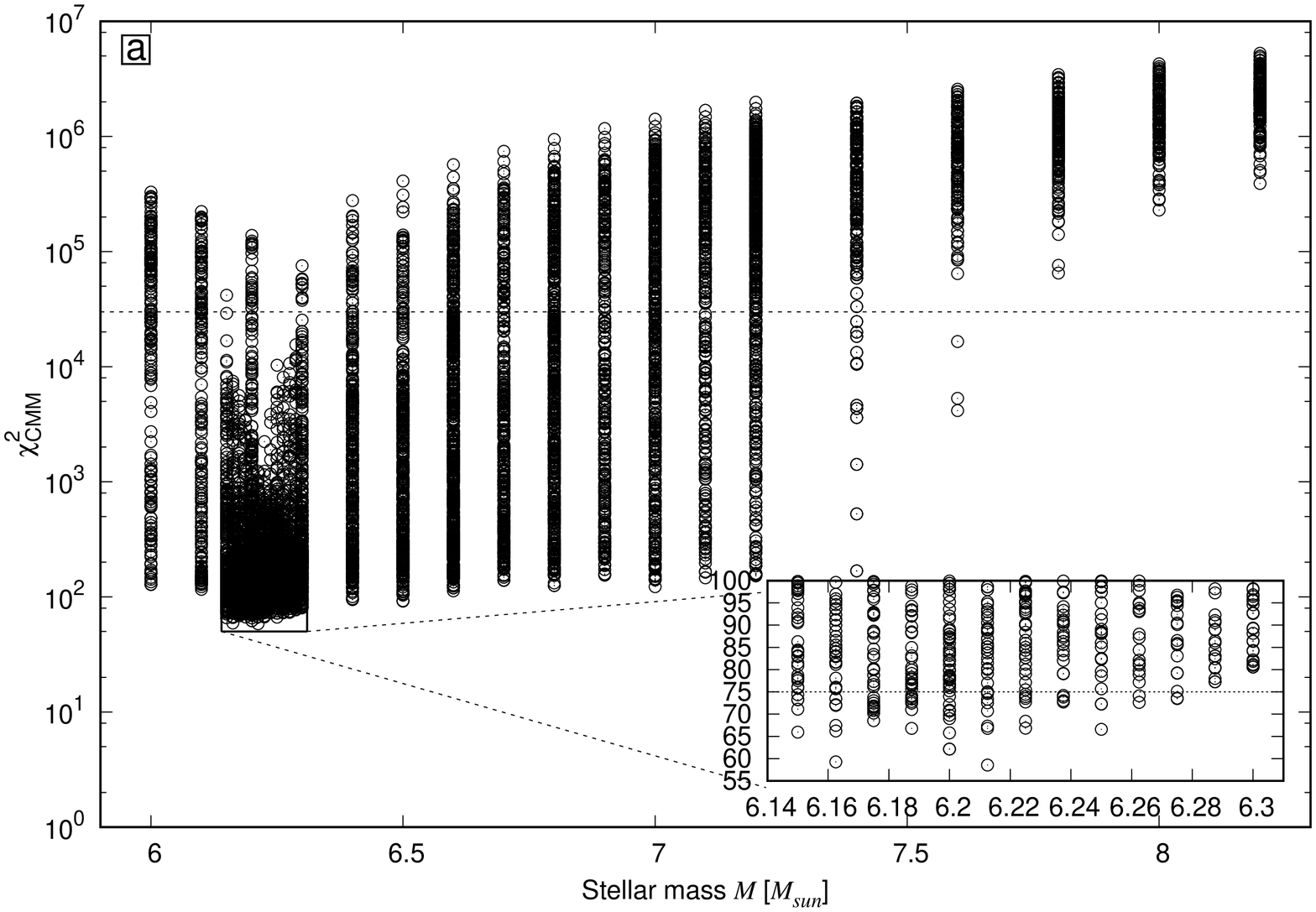}
  \includegraphics[scale=0.3,angle=-90]{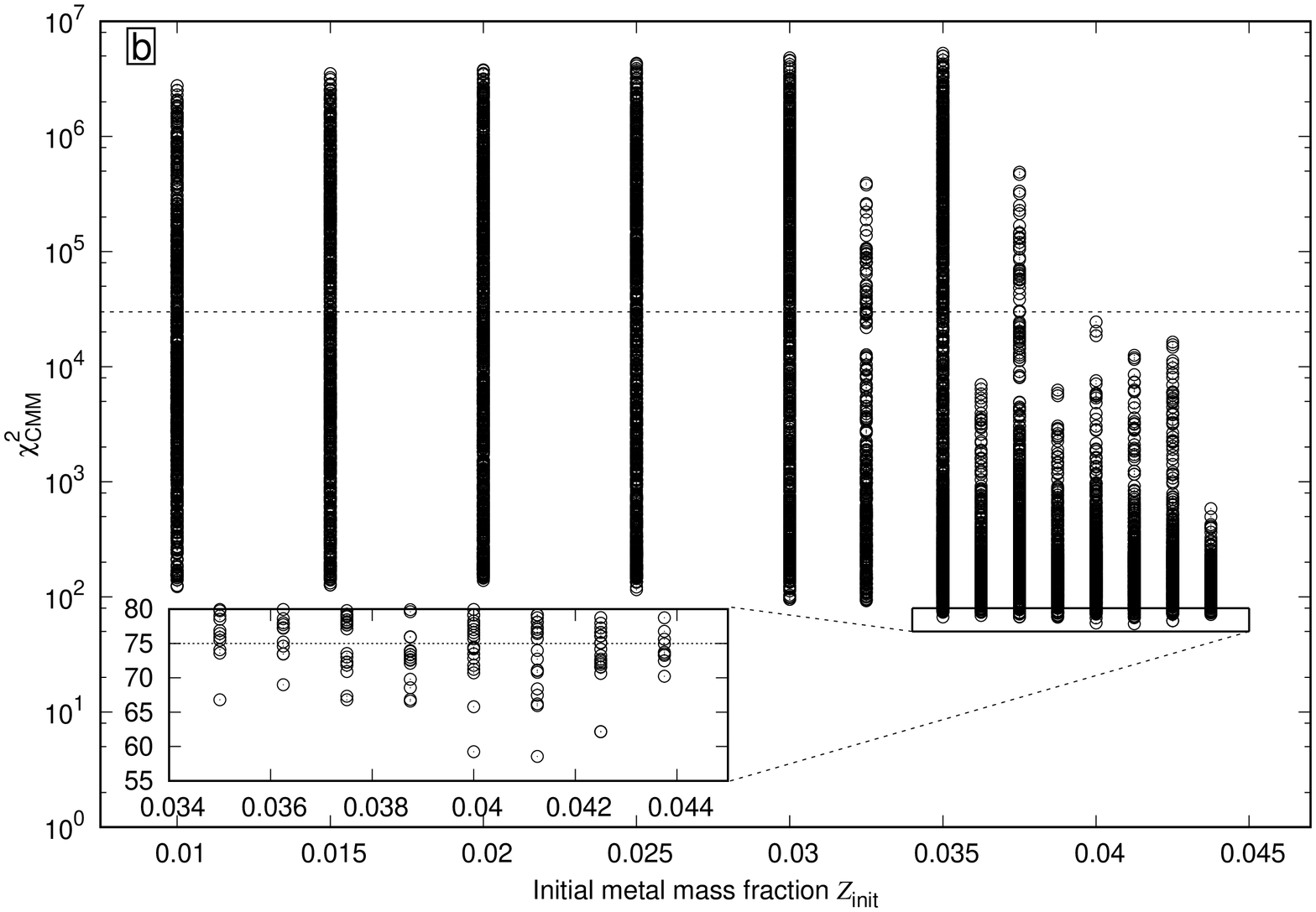}
  \includegraphics[scale=0.3,angle=-90]{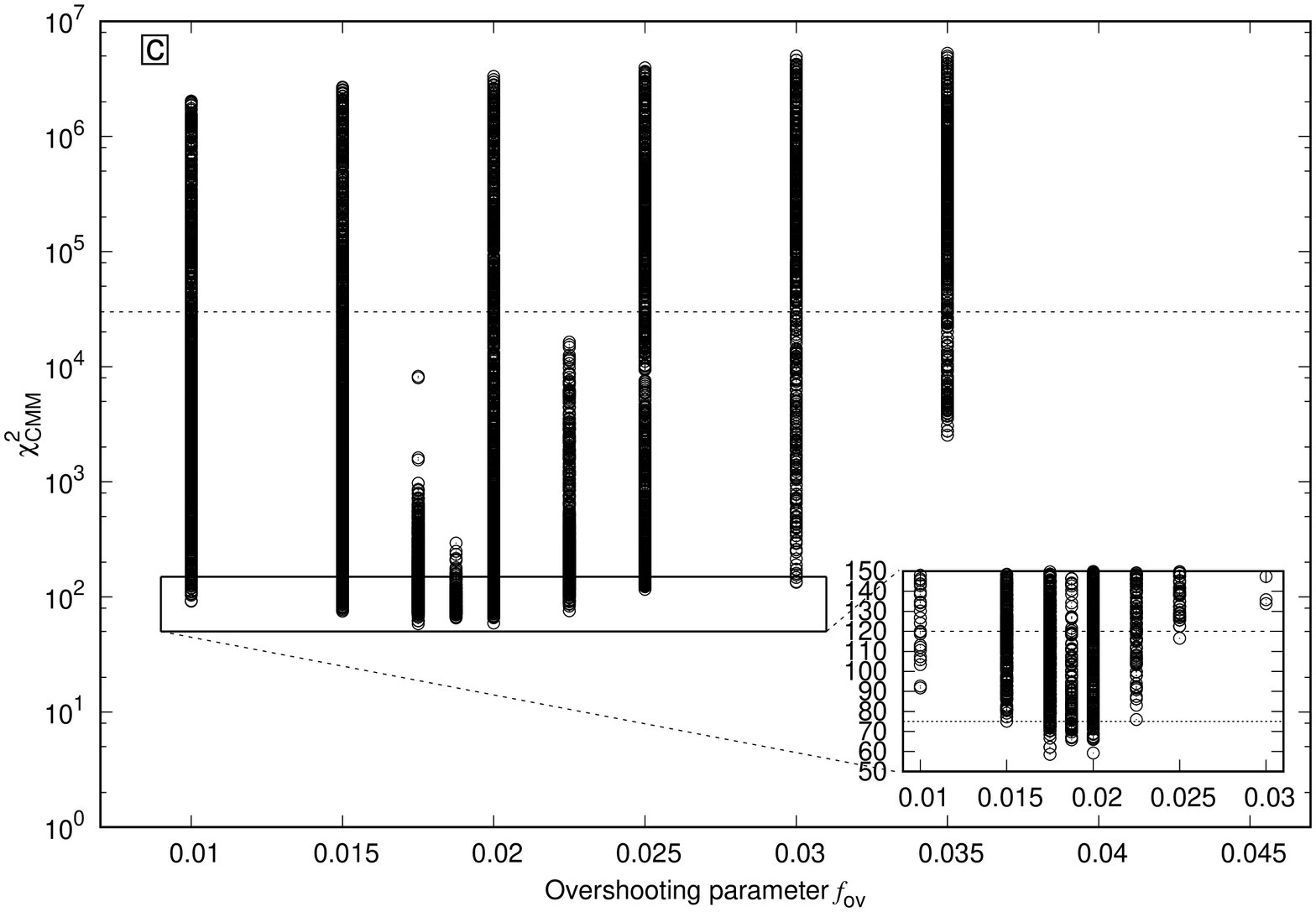}
  \includegraphics[scale=0.3,angle=-90]{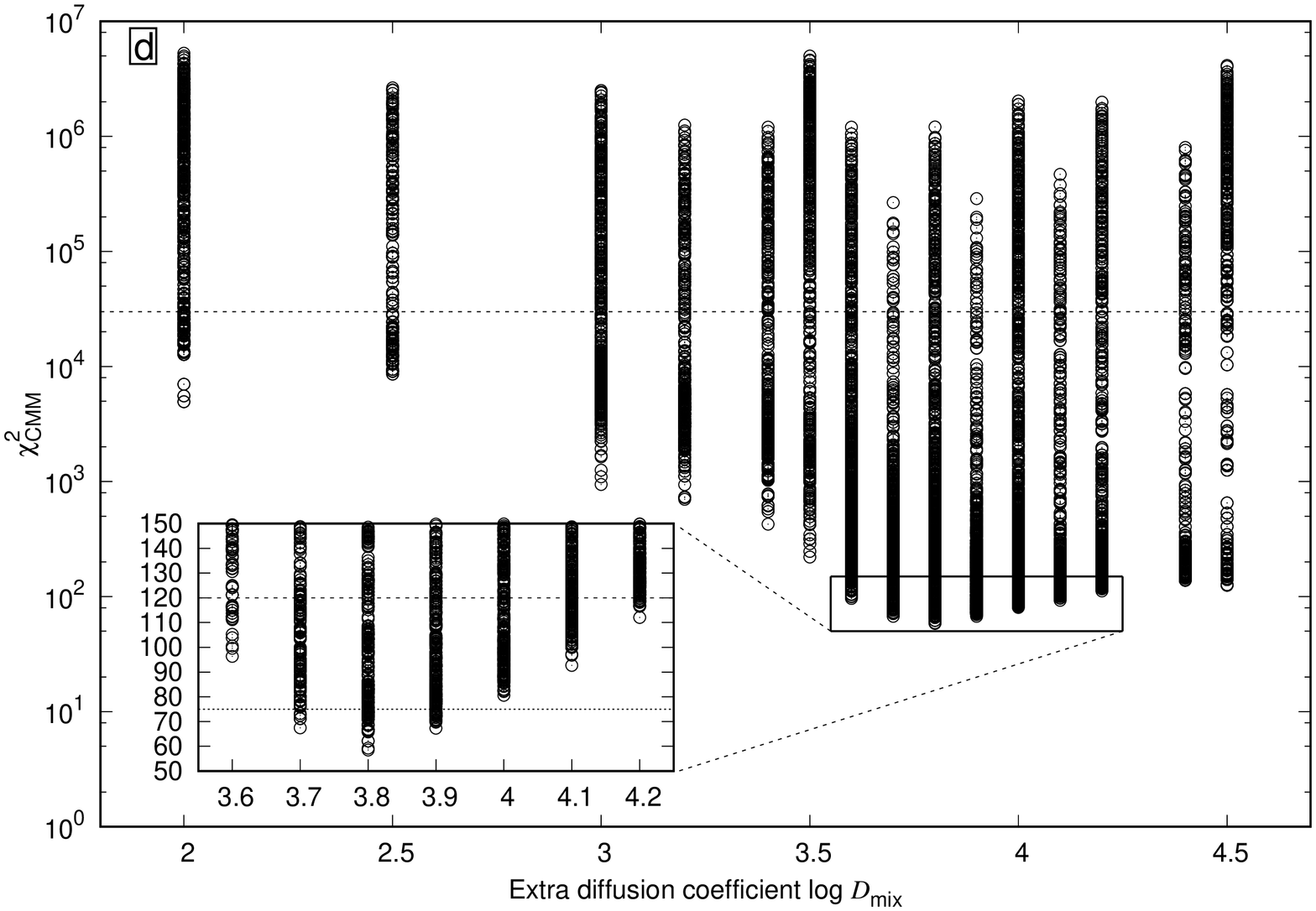}
   \caption{$\chi^2_{\rm CMM}$ as a function of the initial inputs: stellar mass ($M$; Panel (a)), initial metal mass fraction ($Z_{\rm init}$; Panel (b)), overshooting parameter ($f_{\rm ov}$; Panel (c)), and extra mixing coefficient ($\log D_{\rm mix}$; Panel (d)), respectively. The corresponding smaller panels are their zooms which is marked with boxes in those panels.
  }\label{fig.chi.init}
  \end{center}
\end{figure*}

In order to obtain the real best fitting model, the ranges and the steps of the initial input parameter spaces are adjusted step by step during the model calculations. As shown in Table \ref{table_mg}, from Grid A to Grid B to final Grid C, the model grids become denser and denser. Correspondingly, as shown in Figure \ref{fig.chi.init} the minimum value of $\chi^2_{\rm CMM}$ becomes smaller and smaller. Finally, more than 8000 evolutionary tracks are calculated, i.e., more than 8000 CMMs are decided.

It can be seen from Equation \eqref{eq:chiP} and Figure \ref{fig.chi} that there are three ways used to decide the CMMs. The corresponding best fitting models are noted as model MA -- $\chi^2$, MP -- $\chi^2_{P}$, and model MDP -- $\chi^2_{\Delta P}$, respectively. Their fundamental parameters are listed in Table \ref{table_bfm}. For a given evolutionary track, the CMMs decided by different ways might be different, especially for that of $\chi^2_{\Delta P}$, as shown in Figure \ref{fig.chi}. While they are almost consistent overall for constraining the optimal parameter ranges of the target HD 50230 from those calculated models. In the present work, we will mainly analyze the combination term, i.e., $\chi^2$.

\begin{deluxetable}{lccc}
\tablecaption{The parameters of the best fitting models. \label{table_bfm} }
\tablehead{
\multirow{2}{*}{Variables} & \multicolumn{3}{c}{Values} \\
 & \colhead{MA$^{\rm a}$} & \colhead{MP$^{\rm a}$} & \colhead{MDP$^{\rm a}$}
}
\startdata
\multicolumn{4}{c}{initial inputs}\\
\hline

\hline
Mass $M$ (${\rm M_{\odot}}$)           & 6.2125  & 6.1625   & 6.2125 \\
Initial metal abundance $Z_{\rm init}$ & 0.04125 & 0.040 & 0.040 \\
Overshooting parameter $f_{\rm ov}$    & 0.0175 & 0.020 & 0.01875 \\
Extra mixing $\log D_{\rm mix}$        & 3.8  & 3.7  & 3.7 \\
\hline

\hline
\multicolumn{4}{c}{Fundamental parameters}\\
\hline

\hline
Age (Myr)                                       &   61.6 & 63.6 & 62.0 \\
Central hydrogen $X_{\rm C}$                  & 0.306 & 0.306 & 0.297 \\
Effective temperature $T_{\rm eff}$ (K)   &  14923 & 14907 & 14931 \\
Luminosity $L$ (${\rm L_{\odot}}$)                &    1440 & 1443 & 1477\\
Radius $R$ (${\rm R_{\odot}}$)                    & 5.68 & 5.70 & 5.75\\
Surface gravity $\log g$ (c.g.s. unit)          &   3.722 & 3.716 & 3.712  \\
Mass of convective core $M_{\rm cc}$ (${\rm M_{\odot}}$)                                   & 1.028  &  1.025 & 1.021 \\
Radius of convective core $R_{\rm cc}$ (${\rm R_{\odot}}$) &      0.531  & 0.529 & 0.529\\
Period spacing $\Delta\Pi_{l=1}$ (s) & 9051.6 & 9031.1 & 9032.4 \\
$\chi^2$, $\chi^2_{P}$, $\chi^2_{\Delta P}$$^{\rm b}$  & 58.5 & 64.7  & 36.3
\enddata
\tablenotetext{a}{MA, MP, MDP represent the best fitting models which are decided by $\chi^2$, $\chi^2_{P}$, and $\chi^2_{\Delta P}$, respectively.}
\tablenotetext{b}{$\chi^2$, $\chi^2_{P}$, and $\chi^2_{\Delta P}$ correspond to MA, MP, and MDP.}
\end{deluxetable}

\section{Results}\label{sec.result}
Those determined CMMs are shown in Figure \ref{fig.chi.init} for matching goodness $\chi^2_{\rm CMM}$ against different initial inputs: stellar mass ($M$) -- Panel (a), initial metal mass fraction ($Z_{\rm init}$) -- Panel (b), overshooting parameter ($f_{\rm ov}$) -- Panel (c), and extra diffusion mixing coefficient ($\log D_{\rm mix}$) -- Panel (d), respectively. In addition, $\chi^2_{\rm CMM}$ against period spacing $\Delta\Pi_{l=1}$ and the other fundamental parameters, such as central hydrogen $X_{\rm C}$, stellar age $t_{\rm age}$, and radius $R$, are shown in Figures \ref{fig.chi.DP}, \ref{fig.af.FP_all} and \ref{fig.chi.XC}.

\subsection{Buoyancy radius $\Lambda_{0}$}

\begin{figure*}
  \begin{center}
  \includegraphics[scale=0.33,angle=-90]{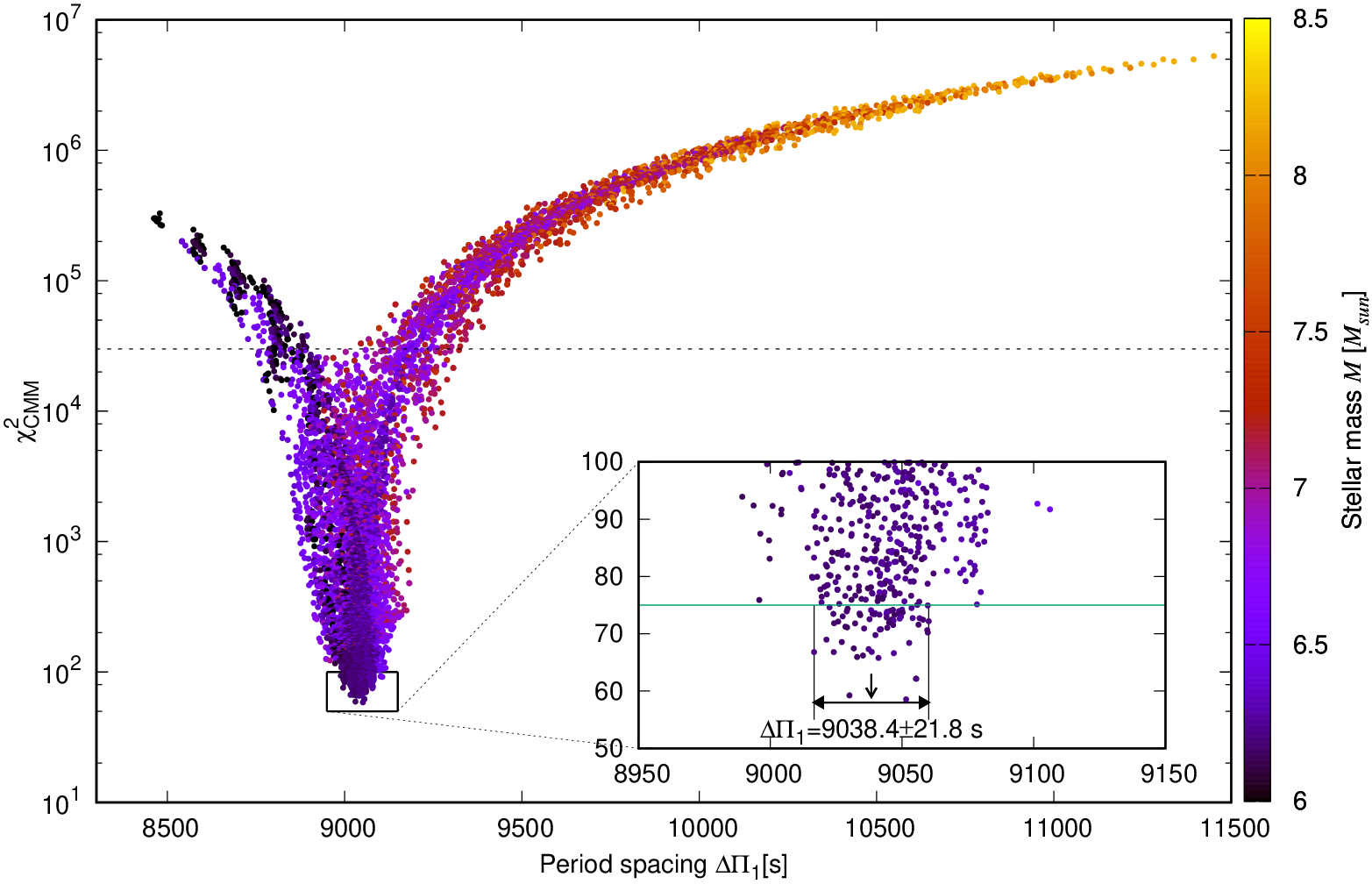}
  \includegraphics[scale=0.33,angle=-90]{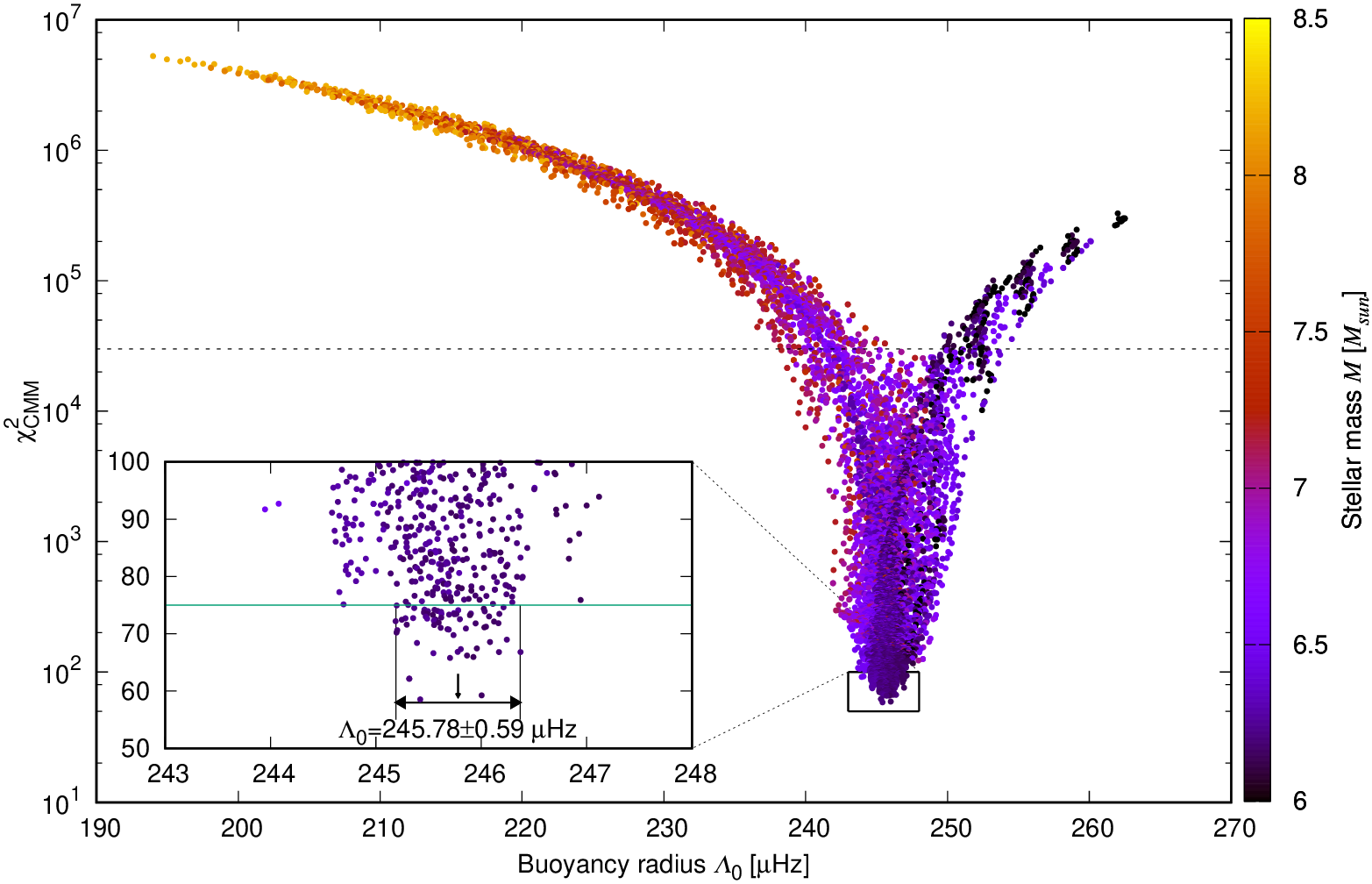}
   \caption{$\chi^2_{\rm CMM}$  as a function of period spacing $\Delta \Pi_{1}$ (right) and of buoyancy radius $\Lambda_{0}$ (left) for all of the calculated CMMs. The smaller panels are zoom boxed on larger panels. The depth of color represents stellar mass ($M_{\odot}$). The green lines in zoom panels represent $\chi^2_{\rm CMM}=75$.
   }\label{fig.chi.DP}
  \end{center}
\end{figure*}

\begin{figure}
  \begin{center}
  \includegraphics[scale=0.34,angle=-90]{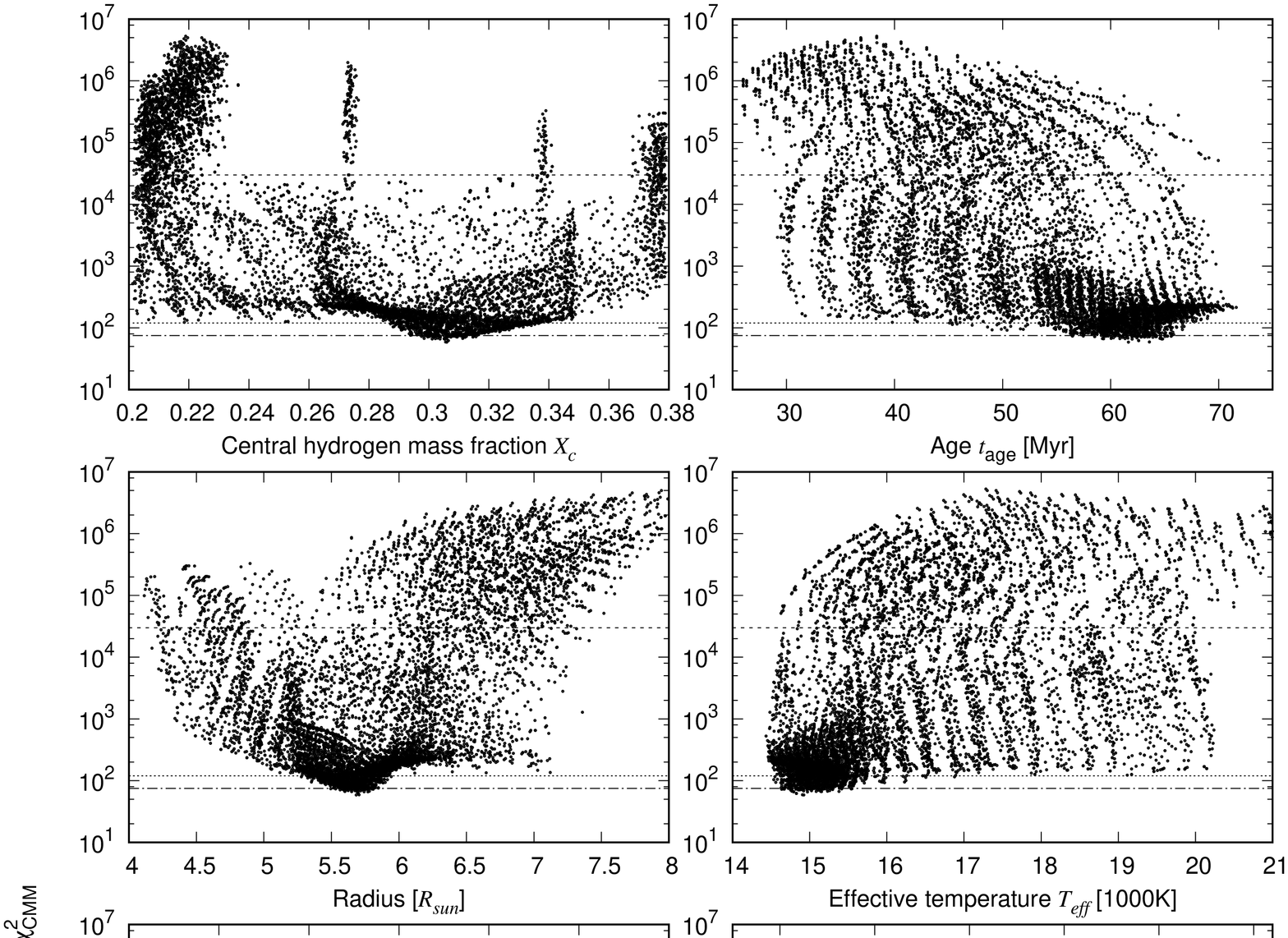}
   \caption{Similar to Figures \ref{fig.chi.DP}, $\chi^2_{\rm CMM}$ as a function of stellar fundamental parameters, central hydrogen ($X_{\rm C}$), age ($t_{\rm age}$), radius ($R$), effective temperature ($T_{\rm eff}$), luminosity ($L$), surface gravity ($\log g$), mass and radius of convective core ($M_{\rm cc}$ and $R_{\rm cc}$), respectively, for all of the calculated CMMs of $X_{\rm init}=0.71$. The horizontal dashed lines represent $\chi^2_{\rm CMM}=120$ and 75, respectively. The zoom panels are shown in Figure \ref{fig.chi.XC}.
  }\label{fig.af.FP_all}
  \end{center}
\end{figure}

The work of \citet[][]{wu2016ApJL} suggested that the acoustic radius $\tau_{0}$ is the only one global parameters that can be accurately measured by the $\chi^2_{\nu}$-matching method between observed frequencies and theoretical model calculating ones for pure p-mode. It means that the pure p-mode mainly carries the information about the acoustic size of p-mode propagation cavity in stars.

Similarly, as shown in Figure \ref{fig.chi.DP}, the CMMs converge into one point (or a narrow range) on the period spacing $\Delta\Pi_{l=1}$ and/or buoyancy radius $\Lambda_{0}$, which characterizes the buoyancy size of the g-mode propagation cavity (more descriptions see Section Appendix \ref{sec.app} --- Propagation velocity of g-mode and buoyancy radius) except for those outliers whose $\chi^2_{\rm CMM}$ are larger than $3\times10^4$.

However, these CMMs distribute into a very large range on other fundamental parameters, as shown in Figure \ref{fig.af.FP_all}, such as the central hydrogen $X_{\rm C}$, radius $R$, surface gravity $\log g$, and the mass of convective core $M_{\rm cc}$. The distribution of CMMs on the radius of convective core $R_{\rm cc}$ has similar behaviors with that of period spacing $\Delta\Pi_{l=1}$ and/or buoyancy radius $\Lambda_{0}$, but it has a slightly larger range (about $R_{\rm cc}\sim0.52-0.54~{\rm R_{\odot}}$).

As shown in Figure \ref{fig.chi.DP}, the distribution of CMMs on period spacing $\Delta\Pi_{l=1}$ and/or buoyancy radius $\Lambda_{0}$ like a flying bird with a pair of asymmetrical wings. Those CMMs can be roughly divided into two parts by a horizontal dashed-line of $\chi^2_{\rm CMM}=3\times10^4$. The normal part which has smaller $\chi^2_{\rm CMM}$ ($<3\times10^4$) construct the `body' of bird, while the outliers which have larger $\chi^2_{\rm CMM}$ ($\geqslant3\times10^4$) have larger or smaller period spacings $\Delta\Pi_{1}$ compared with the normal part and make up the asymmetrical `wings'.

It can be seen from Figure \ref{fig.chi.init} that the centers of the optimal initial inputs are about of 6.2 on $M$, 0.041 on $Z_{\rm init}$, 0.01785 on $f_{\rm ov}$, and 3.8 on $\log D_{\rm mix}$, respectively. Figure \ref{fig.chi.ini.3d} intuitively illustrates the influence of extreme inputs for the value of $\chi^2_{\rm CMM}$. In the figure, point size is proportional to $|\log D_{\rm mix}-3.8|$, i.e., the larger point denotes larger discrepancy from the center value of 3.8 for the initial input $\log D_{\rm mix}$. The light or deep colors represent the discrepancy of $f_{\rm ov}$ from the optimal center value of 0.01875. It can be easily found from Figures \ref{fig.chi.init}, \ref{fig.chi.DP} and \ref{fig.chi.ini.3d} that those outliers possess extreme input parameters in stellar mass $M_{\rm init}$, initial metal mass fraction $Z_{\rm init}$, overshooting parameter $f_{\rm ov}$ and extra diffusion coefficient $\log D_{\rm mix}$.

It can be found from Figures \ref{fig.chi.DP} and \ref{fig.chi.ini.3d} that both of $\Delta\Pi_{1}$ and $\chi^2_{\rm CMM}$ are seriously affected by the extreme initial inputs. The above analysis indicates that the farther initial inputs are far away from the center (or proper) values, the larger values of $\chi^2_{\rm CMM}$ and larger or smaller period spacing $\Delta\Pi_{l=1}$ and/or buoyancy radius $\Lambda_{0}$ the CMMs will have.

As shown in Figure \ref{fig.chi.DP} the period spacing $\Delta\Pi_{1}$ of HD 50230 can be accurately determined from the $\chi^2$-matching method. It is $\Delta\Pi_{1}=9038.4\pm21.8$ s, i.e., $\Delta\Pi_{1}=9016.6-9060.2$ s, which is decided from those 65 selected better candidates whose $\chi^2_{\rm CMM}$ are smaller than 75 as shown in the smaller panel of Figure \ref{fig.chi.DP}. Its relative precision is about 0.24\%. Correspondingly, the buoyancy radius is $\Lambda_{0}=245.78\pm0.59~\mu$Hz, i.e., those selected better candidates distribute in the range of $245.19-246.37$ $\mu$Hz on buoyancy radius $\Lambda_{0}$ as shown in the right panel of Figure \ref{fig.chi.DP}.

What the buoyancy radius $\Lambda_{0}$ is to g-mode oscillations, the acoustic radius $\tau_{0}$ is to p-mode ones (more see Section Appendix \ref{sec.app}). Both of them represent the ``Propagation Time" of oscillation waves of g- and p-modes, respectively, from stellar surface to center, i.e., the ``size" of oscillating cavities. The distribution of CMMs on $\Lambda_{0}$ and initial inputs indicates that proper inputs corresponds suitable oscillating cavities.

\subsection{Fundamental parameters}

\begin{figure}
  \begin{center}
  \includegraphics[scale=0.41,angle=-90]{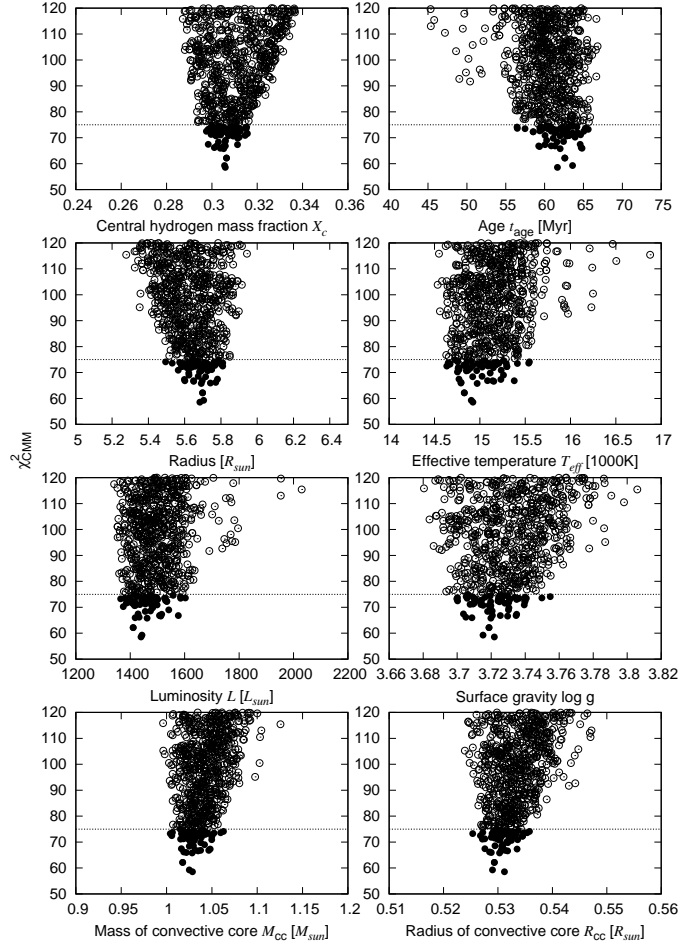}
   \caption{A zoom of Figure \ref{fig.af.FP_all} for those CMMs whose $\chi^2_{\rm CMM}$ is smaller than 120. Horizontal dished-line represent $\chi^2_{\rm CMM}=75$. The 65 selected better candidates are shown in figures with filled points.
  }\label{fig.chi.XC}
  \end{center}
\end{figure}

\begin{figure}
  \begin{center}
  \includegraphics[scale=0.35,angle=-90]{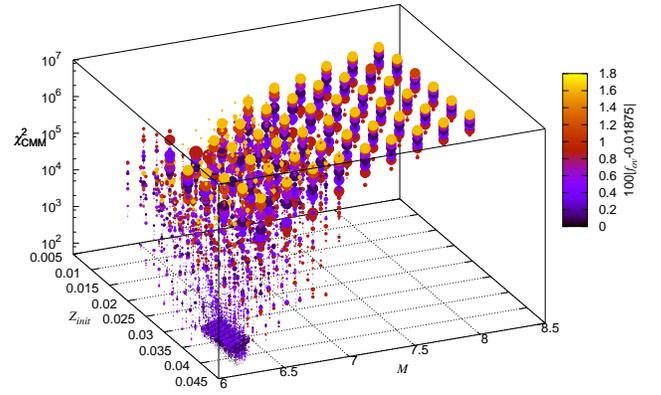}
   \caption{$\chi^2_{\rm CMM}$ as a function of $Z_{\rm init}$ and $M$. The light and deep colors of points are related to $100|f_{\rm ov}-0.01785|$. The size of points is inversely proportional to $|\log D_{\rm mix}-3.8|$.
  }\label{fig.chi.ini.3d}
  \end{center}
\end{figure}

It can be seen from Figure \ref{fig.af.FP_all} that the distribution of CMMs on other fundamental parameters (i.e., $\chi^2_{\rm CMM}$ against the other fundamental parameters) seems disorder compared with $\chi^2_{\rm CMM}$ vs. $\Delta\Pi_{1}$ (Figure \ref{fig.chi.DP}). Therefore, it is very difficult to directly determine the optimal range of the other fundamental parameters from the Figure \ref{fig.af.FP_all} like deciding period spacing $\Delta\Pi_{l=1}$ and buoyancy radius $\Lambda_{0}$ from Figure \ref{fig.chi.DP}. However, the zoom figure (i.e., Figure \ref{fig.chi.XC}) illustrates that those CMMs whose $\chi^2_{\rm CMM}$ are smaller than 120 also converge into a certain relative narrow region on these fundamental parameters, including central hydrogen ($X_{\rm C}$), stellar age ($t_{\rm age}$), radius ($R$), effective temperature ($T_{\rm eff}$), luminosity ($L$), surface gravity ($\log g$), and on the mass and radius of convective core ($M_{\rm cc}$, $R_{\rm cc}$). In Figure \ref{fig.chi.XC}, filled points represent the 65 selected better candidates whose $\chi^2_{\rm CMM}$ are smaller than 75. Based on those better candidates, as shown in Figure \ref{fig.chi.XC}, we obtain the optimal ranges of the fundamental parameters, which are listed in Table \ref{table_range}.

\begin{figure}
  \begin{center}
  \includegraphics[scale=0.4,angle=-90]{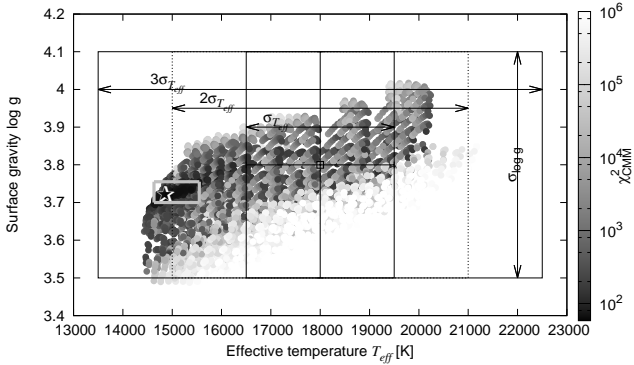}
   \caption{The distribution of CMMs on $\log g$ vs. $T_{\rm eff}$. The solid and dashed thin lines represent the observations and their observational uncertainties: $\sigma_{\log g}$, $\sigma_{T_{\rm eff}}$, $2\sigma_{T_{\rm eff}}$, and $3\sigma_{T_{\rm eff}}$, respectively. They are marked with solid lines with arrows in figure. The gray thick solid box presents the optimal ranges of $T_{\rm eff}$ and $\log g$ which is listed in Table \ref{table_range}. The asterisk symbols model MA.
  }\label{fig.chi.HR}
  \end{center}
\end{figure}

\begin{deluxetable*}{lll}
\tablecaption{The optimal variable range of the fundamental parameters of HD 50230. \label{table_range} }
\tablehead{
 & \multicolumn{2}{c}{Ranges}
 \\ \colhead{Variables} & \colhead{$\chi^2_{\rm CMM}<=75.0^{\rm a}$} & \colhead{68.3\% probability}
}
\startdata
Stellar mass $M$ ($\rm M_{\odot}$)          & $6.15-6.27~(6.21\pm0.06)$     & $6.187\pm0.025^{\rm b}$ \\
Initial metal abundance $Z_{\rm init}$      & $0.034-0.043~(0.041^{+0.002}_{-0.007})$   & $0.0408\pm0.0009^{\rm b}$ \\
Overshooting parameter in core $f_{\rm ov}$ & $0.0175-0.0200$ & $0.0180\pm0.0014^{\rm b}$\\
Extra mixing parameter $\log D_{\rm mix}$   & $3.7-3.9~(3.8\pm0.1)$       & $3.800\pm0.045^{\rm b}$\\
\hline
Period spacing $\Delta\Pi_{1}$ (s)          & $9016.6-9060.2~(9038.4\pm21.8)$ & $9044.75^{+9.35}_{-14.87}$\\
Buoyancy radius $\Lambda_{0}$ ($\mu$Hz)     & $245.19-246.37~(245.78\pm0.59)$ & $245.61^{+0.40}_{-0.25}$\\
\hline
Age (Myr)                                   & $56.5-65.6~(61.6^{+4.0}_{-5.1})$     & $61.72^{+1.89}_{-0.21}$\\
Central hydrogen $X_{\rm C}$                & $0.298-0.316~(0.306^{+0.010}_{-0.008})$   & $0.3058^{+0.0006}_{-0.0007}$ \\
Effective temperature $T_{\rm eff}$ (K)     & $14600-15500~(14900^{+600}_{-300})$   & $14920^{+50}_{-30}$\\
Luminosity $\log L$ (${\rm L_{\odot}}$)     & $3.135-3.205~(3.158^{+0.047}_{-0.023})$   & $3.1585^{+0.0073}_{-0.0054}$\\
Radius $R$ (${\rm R_{\odot}}$)              & $5.50-5.81~(5.68^{+0.13}_{-0.18})$     & $5.689^{+0.018}_{-0.015}$\\
Surface gravity $\log g$ (c.g.s. unit)      & $3.700-3.755~(3.722^{+0.033}_{-0.022})$   & $3.7208^{+0.0020}_{-0.0061}$\\
Mass of convective core $M_{\rm cc}$ (${\rm M_{\odot}}$) & $1.004-1.063~(1.028^{+0.035}_{-0.024})$ &  $1.0276^{+0.0030}_{-0.0039}$\\
Radius of convective core $R_{\rm cc}$ (${\rm R_{\odot}}$)& $0.525-0.536~(0.531^{+0.005}_{-0.006})$ &  $0.5308^{+0.0006}_{-0.0019}$
\enddata
\tablenotetext{a}{The range $x_1-x_2$ represents the minimization and maximization of the parameters for the CMMs whose $\chi^2_{\rm CMM}\leqslant75.0$. For the form of $x^{+dx}_{-dx}$, $x$ corresponds the value of the best fitting model (MA) as shown in Table \ref{table_bfm}. Correspondingly, $+dx$ and $-dx$ express the discrepancy between the $x$ and $x_1$, $x_2$, respectively.}
\tablenotetext{b}{Fitting the probability distribution density with Gaussian function and adopting $1\sigma$. The fitting results are shown in Figure \ref{fig.pd.init}. }
\end{deluxetable*}

As shown in Figure \ref{fig.chi.HR} and Table \ref{table_range} the central hydrogen $X_{\rm C}$ of HD 50230 is about $0.298-0.316$. It indicates that about 55-58\% initial hydrogen has exhausted in stellar center. It is slightly smaller than the result of \citet[][60\%]{Degroote2010Natur}.

According to the radius of the best fitting model MA ($R=5.68~{\rm R_{\odot}}$; see Table \ref{table_bfm}) and the optimal range of radius ($R=5.50-5.81~{\rm R_{\odot}}$; see Table \ref{table_range}), we express the radius of HD 50230 as $R=5.68^{+0.13}_{-0.18}~{\rm R_{\odot}}$. According to Equation \eqref{eq_domega2} and the observation of rotational splitting of p-mode $\Delta f_{\rm obs,p}=0.044\pm0.007~{\rm day}^{-1}$ \citep[][]{Degroote2012AA}, the rotational velocity is $V_{\rm eq}=12.65^{+2.35}_{-2.35}~{\rm km~s}^{-1}$. It is consistent with the spectroscopic observation of $V_{\rm eq}\sin i=6.9\pm1.5~{\rm km~s}^{-1}$ \citep[][]{Degroote2012AA} on the same order of magnitude. Based on these, we can obtain the inclination angle $i$ of the rotating axis to be of $33^{+21}_{-12}{^\circ}$, which is consistent with the prediction of \citet[][$i$ to be $i>20{^\circ}$]{Degroote2012AA}.

As shown in Figure \ref{fig.chi.HR} and Table \ref{table_bfm} the surface gravity of HD 50230 is $\log g=3.722^{+0.033}_{-0.022}$ (for model MA), which is in good agreement with the spectroscopic observation of $\log g=3.8\pm0.3$ \citep[][]{Degroote2012AA}. The effective temperature is about $T_{\rm eff}=14900^{+600}_{-300}$ K), which is consistent with the spectroscopic observation of $T_{\rm eff}=18000\pm1500$ K \citep[][]{Degroote2012AA} with $2\sigma_{T_{\rm eff}}$.

\citet[][]{Degroote2010Natur,Degroote2012AA} and \citet[][]{Szewczuk2014IAUS} suggested that SPB star HD 50230 is a metal-rich ($\log Z/Z_{\odot}=0.3$ dex; $Z_{\odot}=0.02$) star, i.e., $Z\simeq0.04$. In the present work, the optimal range of better candidates in metallicity ($Z_{\rm init}$) is of $0.034-0.043$ (or $Z_{\rm init}=0.041^{+0.002}_{-0.007}$), which is consistent with that of literature.

In addition, the calculations indicates that HD 50230 has a convective core. Its radius ($R_{\rm cc}$) and mass ($M_{\rm cc}$) are $0.525-0.536~{\rm R_{\odot}}$ (or $R_{\rm cc}=0.531^{+0.005}_{-0.006}~{\rm R_{\odot}}$) and $1.004-1.063~{\rm M_{\odot}}$ (or $M_{\rm cc}=1.028^{+0.035}_{-0.024}~{\rm M_{\odot}}$), respectively.

\subsubsection{Statistical Analysis: Probability Distribution of Fundamental Parameters}

\begin{figure*}
  \begin{center}
  \includegraphics[scale=0.48,angle=-90]{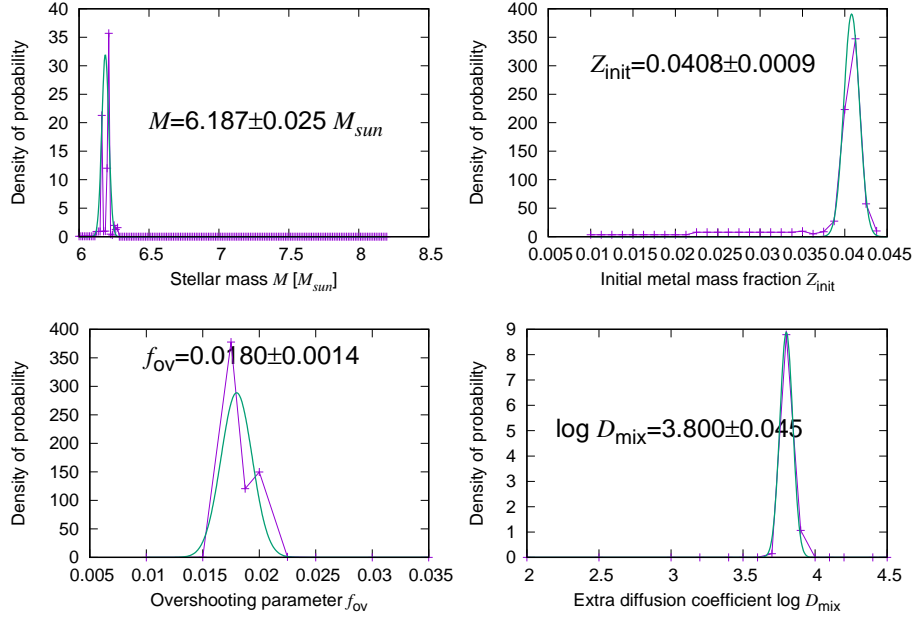}
   \caption{The density of probability as a function of the initial inputs: stellar mass ($M$), initial metal mass fraction ($Z_{\rm init}$), overshooting parameter ($f_{\rm ov}$), and extra mixing coefficient ($\log D_{\rm mix}$), respectively. In the bottom four panels, thick green lines represent Gaussian fit of the density of probability.
  }\label{fig.pd.init}
  \end{center}
\end{figure*}

\begin{figure*}
  \begin{center}
  \includegraphics[scale=0.34,angle=-90]{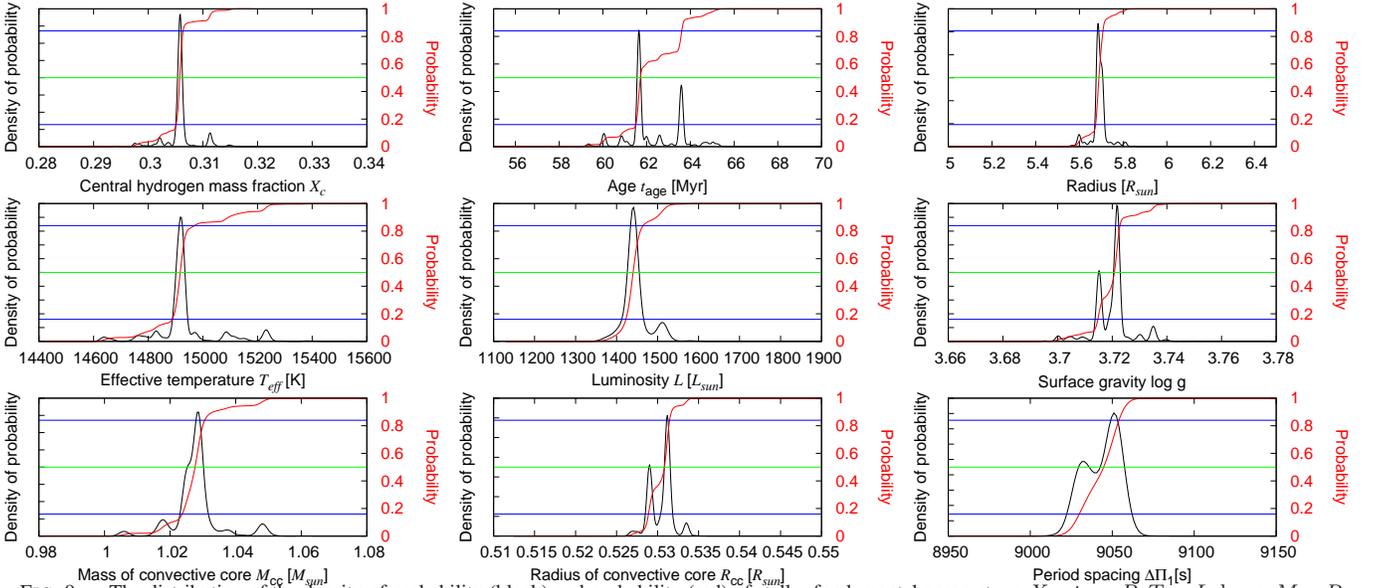}
   \caption{The distribution of the density of probability (black) and probability (red) of stellar fundamental parameters: $X_{\rm C}$, $t_{\rm age}$, $R$, $T_{\rm eff}$, $L$, $\log g$, $M_{\rm cc}$, $R_{\rm cc}$, and $\Delta\Pi_{1}$. The three horizontal lines represent 15.85\% (blue), 50\% (green), and 84.15\% (blue), respectively.
  }\label{fig.pd.par}
  \end{center}
\end{figure*}

Similar to the work of \citet[][]{Giammichele2016ApJS}, \citet[][]{Gai2018ApJ}, and \citet[][]{Tang2018ApJ}, we calculate the Likelihood Function in 5D parameter space,
\begin{equation}
L(x_{1}, ~x_{2},~x_{3},~x_{4},~x_{5}) \propto e^{-\frac{1}{2}\chi^2},
\end{equation}
from the merit function, $\chi^2(x_{1}, ~x_{2},~x_{3},~x_{4},~x_{5})$. For the five independent variables, four of them are the initial input parameters, i.e., stellar mass ($M$), initial metal mass fraction ($Z_{\rm init}$), convective overshooting parameter ($f_{\rm ov}$) in center, and the extra diffusion coefficient ($\log D_{\rm mix}$). The another one represents the evolutionary status of star, such as stellar age ($t_{\rm age}$), radius ($R$), center hydrogen mass fraction ($X_{\rm C}$). For a chosen parameter, such as $x_{1}$, its density of probability function can be defined from the following integration over the full parameter range,
\begin{equation}
\begin{split}
p(x_{1}) & \propto \int L(x_{1}, ~x_{2},~x_{3},~x_{4},~x_{5})dx_{2}dx_{3}dx_{4}dx_{5} \\ & \propto \int e^{-\frac{1}{2}\chi^2}dx_{2}dx_{3}dx_{4}dx_{5} .
\end{split}
\end{equation}
And then, normalizing the density of probability function, i.e., assuming that the integration of $p(x_{1})dx_{1}$ over the allowed parameter range $[x_{1,\rm min},~x_{1,\rm max}]$ is equal to 1, to decide the normalization factor
\begin{equation}
\int p(x_{1})dx_{1} = 1.
\end{equation}
Based on such statistical analysis method, the distributions of the density of probability function for initial input parameters and the other stellar fundamental parameters are determined and shown in Figures \ref{fig.pd.init} and \ref{fig.pd.par}, respectively.

As shown in Figure \ref{fig.pd.init} the profile of the density of probability function is similar to a $\delta-$like function. The probability centers into a narrow range. In order to determine their optimal values and the corresponding uncertainties, we adopt Gaussian function to fit the density of probability for the four initial input parameters: $M$, $Z_{\rm init}$, $f_{\rm ov}$, and $\log D_{\rm mix}$, because the parameter space of ($M$, $Z_{\rm innit}$, $f_{\rm ov}$, $\log D_{\rm mix}$) has lower resolution. However, as shown in Figure \ref{fig.pd.par}, we through integrating the density of probability to determine them for other parameters. The final decided optimal values are listed in Table \ref{table_range}.

It can be seen from Table \ref{table_range} and Figures \ref{fig.pd.init} and \ref{fig.pd.par} that compared to these values, which are decided by these CMMs whose $\chi^2_{\rm CMM}$ are smaller than 75, the statistical analyses have lower uncertainties. In the present work, we adopt the former method to determine the optimal range of the fundamental parameters.

\subsection{Asteroseismic analysis}

\begin{figure}
  \begin{center}
  \includegraphics[scale=0.4,angle=-90]{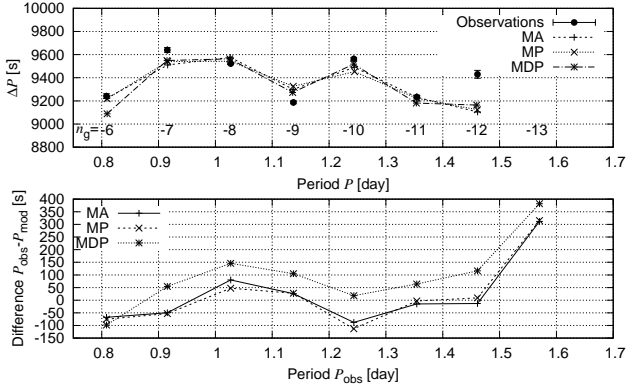}
   \caption{Period spacings $\Delta P$ as a function of periods $P$ --- upper panel; the differences of periods between observations and the best fitting models $P_{\rm obs}-P_{\rm mod}$ as a function of periods $P_{\rm obs}$ --- bottom panel. MA, MP, and MDP represent different best fitting models which are listed in Table \ref{table_bfm}. The radial orders $n_{\rm g}$ are shown in figure with text.
  }\label{fig.DP}
  \end{center}
\end{figure}

The period spacings ($\Delta P$) of the observations and the bet fitting models and the differences of periods between them ($P_{\rm obs}-P_{\rm mod}$) are shown in Figure \ref{fig.DP}.

It can be found from the upper panel of Figure \ref{fig.DP} that the period spacings $\Delta P$ of the best fitting models are almost consistent with those of observations except for the last one ($g_{13}$). As shown in bottom panel of Figure \ref{fig.DP} the observed periods are larger than those of model MDP, which is decided by matching period spacing $\Delta P_{i}$ between observations and models except for the first mode ($g_{6}$). For models MA, which is decided by matching both of period $P_{i}$ and period spacing $\Delta P_{i}$, and MP, which is decided by matching period $P_{i}$, their periods are consistent with observations within about 100 s except for the last one ($g_{13}$) whose period is smaller than the observed one to be about 300 s.

It can be found from the upper panel of Figure \ref{fig.DP} that the observed period spacings and the model calculated ones have different tendencies on the end of larger period. The period spacing decreases with the increase of period for models when $P\gtrsim1.3$ days. Based on those best fitting models, the last observed period ($g_{13}$) seems to be an outlier for the sequence of ($l=1$, $m=0$).

\begin{figure}
  \begin{center}
  \includegraphics[scale=0.47,angle=-90]{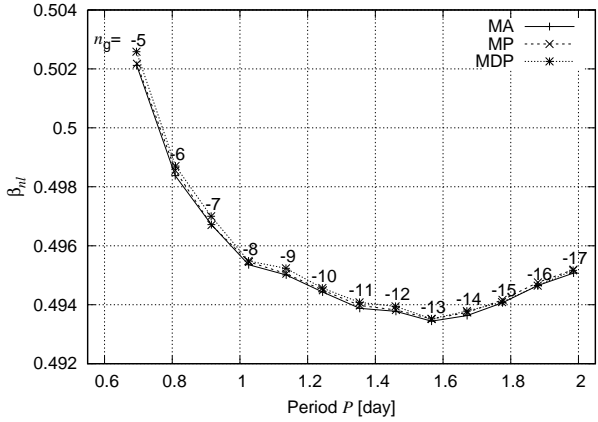}
   \caption{Rotational splitting parameter $\beta_{nl}$ as a function of period $P$ for the three best fitting models: MA, MP, and MDP.
  }\label{fig.bets}
  \end{center}
\end{figure}

The period difference $\delta P\sim300$ s in the last mode ($P_{n_{\rm g}=-13}=135359.94$ s) corresponds to about $\delta f\thicksim-0.0014~{\rm day}^{-1}$ (or $-0.016~\mu$Hz) in frequency. As shown in Figure \ref{fig.bets} the rotational splitting parameters $\beta_{nl}$ are around 0.5 for all of these modes. As a matter of fact, for all of the calculated CMMs, $\beta_{nl}$ are around 0.5. They are pure g-mode. For the ``outlier" corresponding mode ($g_{13}$), its rotational parameter $\beta_{nl}$ is about 0.4934. Therefore, the frequency difference of $\delta f\thicksim-0.0014~{\rm day}^{-1}$ might be explained as a rotation splitting of $\Delta f^{I}_{{\rm obs,g},l=1,m=-1}\thicksim-0.0014~{\rm day}^{-1}$, which corresponds a rotational frequency of $\Omega_{\rm rot,s}\thicksim0.0028~{\rm day}^{-1}$. Surely such rotational frequency is far smaller than that of observation $\Omega_{\rm rot,s}\backsimeq0.044 ~{\rm day}^{-1}$ ($=0.509~\mu$Hz, see Section \ref{sec.HD5}), which is calculated from the rotational splitting of p-mode. Therefore, it is not suitable that the period difference of last mode ($g_{13}$) between observations and best fitting models is explained as rotational splitting.

\begin{figure}
  \begin{center}
  \includegraphics[scale=0.42,angle=-90]{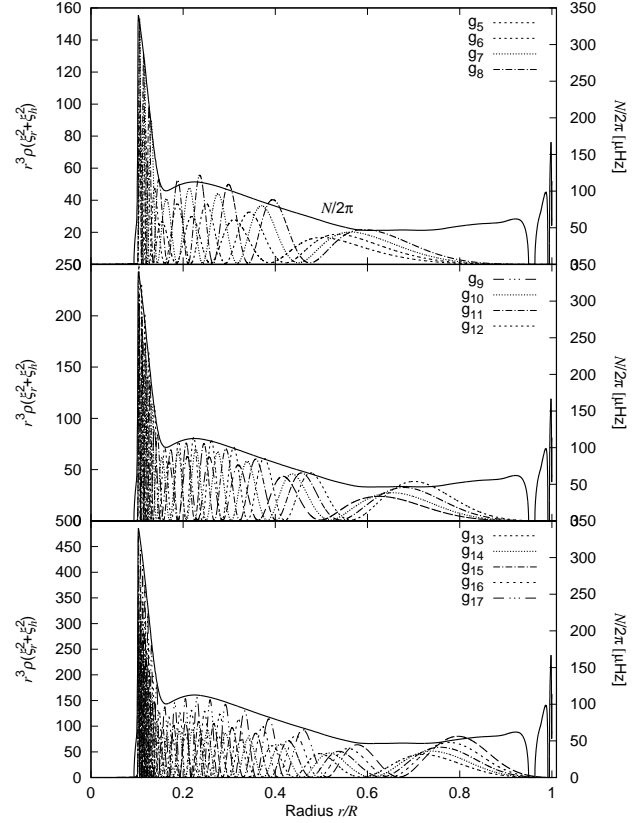}
   \setlength{\abovecaptionskip}{17pt}
   \caption{The right and left axes represent Buoyancy frequency $N/2\pi$ and $r^3\rho(\xi^2_{r}+\xi^2_{h})$ of model MA, which presents the distribution of oscillation energy, respectively. $g_{5}$, $g_{6}$, ..., and $g_{17}$ correspond to the radial orders $n_{\rm g}=-5,~-6,~...,~{\rm and} ~-17$, which are shown in Figure \ref{fig.bets}.
  }\label{fig.N.and.energ}
  \end{center}
\end{figure}

As shown in Figure \ref{fig.N.and.energ} the most of the oscillation energy are trapped in the $\mu$-gradient region, which is nearby the convective core, for the best fitting model: model MA. The residual oscillation energy distribute into the outer region of $\mu$-gradient region but within $r/R\lesssim0.85$ for low-order modes. For high-order modes (see bottom panel of Figure \ref{fig.N.and.energ}), the oscillation energy partly extends to stellar surface. According to the theory of stellar oscillations, therefore, as shown in Figure \ref{fig.DP}, the period spacings $\Delta P$ will decreases with the increase of period $P$ for high-order g-mode since they have larger propagation cavity compared with the low-order modes.

For the period discrepancy of the last mode ($g_{13}$), based on the calculated theoretical models in the present work there are two possibilities might explain it. The first one: it is really an outlier. It might not belong the prograde mode sequence of ($l=1$, $m=0$).
The second one: it is caused by certain possible physical mechanism which can partly decrease the frequency of the last mode to make it consistent with the observation, such as the mode trapping. But, it is not properly considered in the present theoretical models.

It can be seen from Figure \ref{fig.DP} that there is a sine-like signal on the period discrepancy between the observations and the best fitting models. The similar phenomena also appear in the previous works, such as \citet[][Figures 4, 6, A.1, and A.2]{Moravveji2015AA}, \citet[][Figure 8]{Moravveji2016ApJ}, and \citet[][Table 3]{Buysschaert2018arXiv}. Such sine-like signal might be caused by a thin layer in stellar interior \citep[see e.g.,][]{Gough1990MNRAS,jcd2003}, which is not considered or improperly expressed in the present theoretical models. For the best fitting model of HD 50230, the strength of mode trapping in $\mu$-gradient region beyond the convective core might be a possible factor. On the other hands, perhaps, the present best fitting model need an extra mode trapping cavity to slightly revise the periods. We will investigate this question in depth in future.

\subsection{The influences: outlier, initial hydrogen, and observational uncertainties}

\begin{figure}
  \begin{center}
  \includegraphics[scale=0.4,angle=-90]{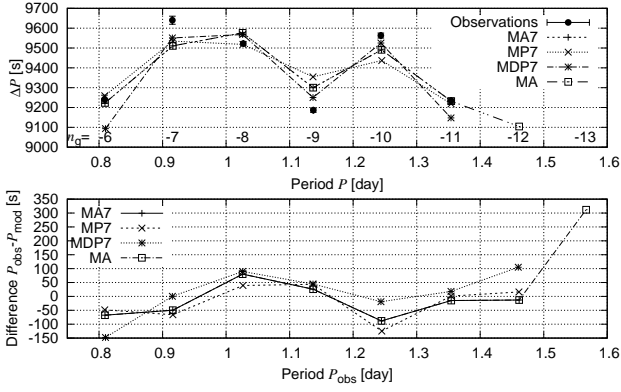}
   \caption{Similar to Figure \ref{fig.DP}, but for a normal case. Here, the outlier (mode $g_{13}$) is not considered for constraining the theoretical models. We only use the former 7 periods to constrain theoretical models. The corresponding best fitting models are noted as models MA7, MP7, and MDP7, respectively.
  }\label{fig.DP6}
  \end{center}
\end{figure}
\begin{figure}
  \begin{center}
  \includegraphics[scale=0.4,angle=-90]{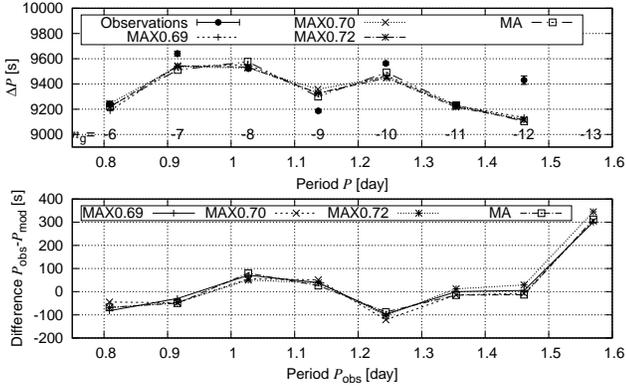}
   \caption{Similar to Figures \ref{fig.DP} and \ref{fig.DP6}, but for the best fitting models of initial hydrogen $X_{\rm init}=0.69$, $0.70$, $0.71$ and $0.72$, respectively. They are noted as models MAX0.69, MAX0.70, MA, and MAX0.72, respectively, compared to the best fitting model: model MA whose initial hydrogen is 0.71.
  }\label{fig.DPX}
  \end{center}
\end{figure}

\begin{figure}
  \begin{center}
  \includegraphics[scale=0.4,angle=-90]{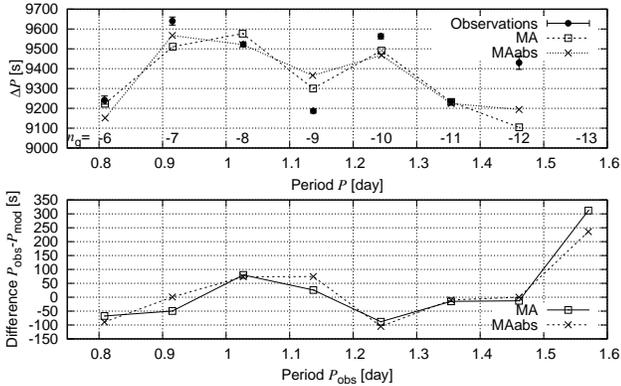}
   \caption{Similar to Figures \ref{fig.DP}, \ref{fig.DP6}, and \ref{fig.DPX}, but for the best fitting models of model MA and model MAabs.
  }\label{fig.DPabs}
  \end{center}
\end{figure}

In the above sections, we used all of the 8 modes to constrain theoretical models which are calculated with a fixed initial hydrogen $X_{\rm init}$. In the process, we adopt the observational uncertainties as weight factors. In the section, we will discuss the influences of different initial hydrogen, observational uncertainties and the last mode for constraining the theoretical models and deciding the final best fitting models.

Firstly, similar to the above, but we only use the former 7 modes (i.e., $g_{6}$, $g_{7}$, ..., and $g_{12}$) to constrain theoretical models. The last mode ($g_{13}$) is not considered. Finally, the corresponding best fitting models are noted as models MA7, MP7, and MDP7, respectively. The period spacings and period differences between observations are shown in Figure \ref{fig.DP6}. It can be seen from Figures \ref{fig.DP} and \ref{fig.DP6} that the final results are almost fully the same for with or without the last mode ($g_{13}$) in observations. It indicates that eliminating the last mode from observations merely partly decrease the value of $\chi^2_{\rm CMM}$ of the best fitting model from 58.5 (model MA) to 53.5 (model MA7). In fact that model MA and model MA7 are the same model. It means that the last mode ($g_{13}$; ``outlier") do not effectively work for constraining the theoretical models in the present work.

Secondly, we change the initial hydrogen $X_{\rm init}$ from the fixed value 0.71 to 0.69, 0.70, and to 0.72 and calculate the corresponding theoretical models to decide best fitting model. For the added calculations, their parameter resolutions are same with the above calculations (see Table \ref{table_mg}). Finally, more than 5500 evolutionary tracks are calculated for the added initial hydrogen $X_{\rm init}=0.69$, $0.70$, and $0.72$. Similar to the best fitting model of model MA, we notes the corresponding best fitting models as models MAX0.69, MAX0.70, and MAX0.72, respectively. Their $\chi^2_{\rm CMM}$ are about 62.0, 65.8, and 60.0, respectively. They are larger than that of model MA (58.5).

It can be seen from Figure \ref{fig.DPX} that the best fitting model does not obviously become better or worse when slightly change (increase or decrease) the initial hydrogen $X_{\rm init}$. The periods of the best fitting models almost overlap each other. On the other hands, their $\chi^2_{\rm CMM}$ are on the same levels. They are close to about 60.

It can be seen from Table \ref{table_obs} (the observations) that the observational uncertainties of periods range from 5.5 s ($g_{8}$) to about 32 s ($g_{13}$). The ratio between them is about 5.8. Correspondingly, in Equation \ref{eq:chiP}, their weight ratio is up to about 34 for calculating the value of $\chi^2$ between the two modes.

In order to analyze the influence of the observational uncertainties for the final best fitting model, we change the $\chi^2$-matching formulas (i.e., Equation \eqref{eq:chiP}) as:
\begin{equation}\label{eq:absP1}
\begin{split}
\Delta_{P}=&\frac{1}{N}\sum^{N}_{i=1}| P^{\rm obs}_{i}-P^{\rm mod}_{i}|,\\
\Delta_{\Delta P}=&\frac{1}{N-1}\sum^{N-1}_{i=1}|\Delta P^{\rm obs}_{i}-\Delta P^{\rm mod}_{i}|,\\
\Delta_{\rm all}=&\frac{1}{2N-1}\left[N\Delta_{P}+(N-1)\Delta_{\Delta P} \right],
\end{split}
\end{equation}
to constrain theoretical models. Here, $\Delta_{P}$, $\Delta_{\Delta P}$, and $\Delta_{\rm all}$ represent the means of differences between observations and theoretical models for period $P$, period spacing $\Delta P$, and both of them, respectively. Correspondingly, the $\Delta_{\rm all}$-minimum model denotes model MAabs whose $\Delta_{\rm all}$ is about 84.9 s. For model MA, the corresponding $\Delta_{\rm all}$ is about 91.1 s. The difference of $\Delta_{\rm all}$ between the two models is mainly contributed by the last period and the last period spacing since they have lower weight compared with the other modes. The period discrepancies with observations of mode $g_{13}$ decrease from about 326 s for model MA to about 235 s for model MAabs as shown in Figure \ref{fig.DPabs}.

\begin{figure}
  \begin{center}
  \includegraphics[scale=0.4,angle=-90]{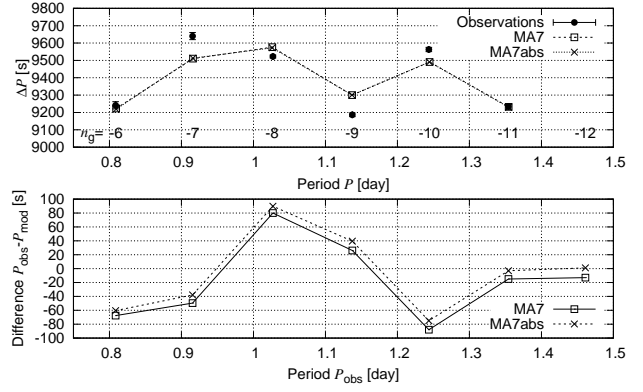}
   \caption{Similar to Figures \ref{fig.DP}, \ref{fig.DP6}, \ref{fig.DPX}, and \ref{fig.DPabs}, but for the best fitting models of model MA7 and model MA7abs.
  }\label{fig.DPabs2}
  \end{center}
\end{figure}
The initial parameters of model MA and model MAabs are the same, except for MAabs has slightly higher initial metal abundance ($Z_{\rm init}=0.04375$). In addition, their fundamental parameters are almost consistent with each other. As shown in Figure \ref{fig.DPabs} the periods and period spacings for the best fitting model (model MAabs) are not changed significantly. It indicates that the final results are not seriously changed whether using the observational uncertainties as a weight or not for determining the best fitting model.

In addition, similarly, we eliminate the last mode ($g_{13}$) from the observations and adopt Equation \eqref{eq:absP1} to determine the best fitting model which is named model MA7abs similar to model MA7. The $\Delta_{\rm all}$ of model MA7abs is 53.3 s. Correspondingly, that of model MA7 is about 56.0 s. For models MA7 and MA7abs, they are located on a common evolutionary track, i.e., their initial parameters are the same. In addition, they are two adjacent models, i.e., their fundamental parameters are merely different.

As shown in upper panel of Figure \ref{fig.DPabs2} the period spacings overlap with that of model MA7. For period differences $\delta P=P_{\rm obs}-P_{\rm mod}$, model MA7abs is slightly larger than model MA7 overall. However, the mean values of $|\delta P|$ are 43.8 and 48.5 s for models MA7abs and MA7, respectively.

It can be found from Figures \ref{fig.DP6}, \ref{fig.DPabs}, and \ref{fig.DPabs2} that the final results almost are not affected whether using the observational uncertainties and the last mode ($g_{13}$) to constrain theoretical models or not for the present calculated models.

\section{{\bf Summary }}\label{sec.con.dis}

SPB star HD 50230 is the primary component of a binary system. It has observed about 137 days with CoRoT satellite and more than 560 frequencies are extracted. There are 8 modes to be identified as likely low-order g-mode with $l=1$ and $m=0$ among the extracted modes due to almost uniform period spacings among them. In addition, the period spacings periodically vary with periods. In the present work, we make model calculations and analyze the 8 modes with high-precision asteroseismology. Finally, the investigation can be briefly concluded as follows:

\rmnum{1}: Similar to pure p-mode oscillations--- the oscillation frequencies mainly carries the information of the size of oscillation wave propagation cavity and the acoustic radius $\tau_{0}$ is the only global parameter that can be precisely measured by the $\chi^2$-matching method between observed frequencies and model calculations \citep[][]{wu2016ApJL}, {\bf the buoyancy radius $\Lambda_{0}$ also can be easily and precisely measured with similar method for pure g-mode oscillations ($m=0$), as shown in Figure \ref{fig.chi.DP}, compared to the other parameters, such as stellar age, effective temperature, and radius, which are shown in Figure \ref{fig.af.FP_all}. This is because that the distribution of CMM on buoyancy radius is not sensitive for initial input parameters compared to the other fundamental parameters.} Both of acoustic radius and buoyancy radius represent the ``Propagation time" of oscillation waves from stellar surface to center for pure p- and g-mode, respectively.

\rmnum{2}: Based on the calculated models, we find that the value of $\chi^2_{\rm CMM}$ and the distribution of CMMs on buoyancy radius $\Lambda_{0}$ can be slightly affected by some extreme initial inputs. Finally, we obtain that the buoyancy radius of HD 50230 is of $\Lambda_{0}=245.78\pm0.59~\mu$Hz with a higher relative precision of 0.24\%. Correspondingly, the period spacing of HD 50230 is $\Delta\Pi_{l=1}=9038.4\pm21.8$ s.

\rmnum{3}: HD 50230 is a metal-rich ($Z=0.041^{+0.002}_{-0.007}$) moderate massive star with a mass of $M=6.21\pm0.06~{\rm M_{\odot}}$ and located on the middle phase of the main-sequence branch with an age of $t_{\rm age}=61.6^{+4.0}_{-5.1}$ Myr. About 57\% initial hydrogen are exhausted in its center ($X_{\rm C}=0.306^{+0.010}_{-0.008}$), which is close to \citet[][]{Degroote2010Natur} estimated (60\%). In addition, HD 50230 has a convective core with a radius of $R_{\rm cc}=0.531^{+0.005}_{-0.006}~{\rm R_{\odot}}$. The corresponding convective core mass is $M_{\rm cc}=1.028^{+0.035}_{-0.024}~{\rm M_{\odot}}$.

\rmnum{4}: Based on the optimal range of stellar radius $R=5.50-5.81~{\rm R_{\odot}}$, we obtain that the rotational velocity of $V_{\rm eq}=12.65^{+2.35}_{-2.35}$ km s$^{-1}$ with an inclination angle of $i=33^{+21}_{-12}{^\circ}$.

\rmnum{5}: In order to interpret the structure in the observed period spacing pattern of HD 50230, the exponentially decaying diffusive core overshooting ($f_{\rm ov}=0.0175-0.0200$) and the extra diffusive mixing ($\log D_{\rm mix}=3.7-3.9$) should be taken into account in theoretical models.

\rmnum{6}: The theoretical models indicate that, for the 8 modes, at least 7 modes can be well explained as dipole g-modes of $(l,~m)=(1,~0)$. Their period discrepancies between the observations and the best fitting models are within 100 s. For the last mode ($g_{13}$), it is almost up to 300 s. In the present work, we still do not find a suitable interpretation for such large discrepancy, but we exclude the possibility that the discrepancy is caused by rotational splitting, i.e., it is not the mode of $(l,~m)=(1,~-1)$.

In the present work, the rotational effects are not taken into account in theoretical models. \citet[][]{Degroote2012AA} predicted the rotational effects should be considered when interpreting the structure in the observed period spacing pattern. Since it will slightly change the period of g-mode \citep[][]{Aerts&Dupret2012}. Perhaps, it is helpful in explaining the larger period discrepancy between observation and best fitting model. We will consider it in the next work.

\acknowledgments
This work is co-sponsored by the NSFC of China (Grant Nos. 11333006, 11503076, 11503079, 11773064, 11873084, and 11521303), and by Yunnan Applied Basic Research Projects (Grant No. 2017B008). The authors gratefully acknowledge the computing time granted by the Yunnan Observatories, and provided on the facilities at the Yunnan Observatories Supercomputing Platform. The authors also express their sincere thanks to Pro. Guo, J.H., Dr. Zhang, Q.S., Dr. Su, J., and Dr. Chen, X.H. for their productive advices.  And finally, the authors are cordially grateful to an anonymous referee for instructive advice and productive suggestions to improve this paper overall.

\appendix
\section{Appendix A: Propagation Velocity of g-mode and buoyancy radius}\label{sec.app}
The propagation velocity of p-mode (i.e., adiabatic sound speed $c$) and the corresponding acoustic radius ($\tau_{0}$) are shown in the works of \citet[][]{Unno1989nosbook}, \citet[][]{jcd2003} and \citet[][]{Aerts2010} in detail. Here, we retrospect it briefly as the background of those of g-mode as following.

For high radial order oscillations, the oscillation equation can be approximately expressed as following in the Cowling approximation \citep[][Eq. (5.17) in page 76]{jcd2003},
\begin{equation}\label{eq:oseq}
\frac{{\rm d}^2\xi_{r}}{{\rm d}r^2}=\frac{\omega^2}{c^2}\left(1-\frac{N^2}{\omega^2}\right)\left(\frac{S_{l}^2}{\omega^2}-1\right)\xi_{r},
\end{equation}
or
\begin{equation}\label{eq:oseqk}
\frac{{\rm d}^2\xi_{r}}{{\rm d}r^2}=-K(r)\xi_{r},
\end{equation}
where
\begin{equation}\label{eq:Kr}
K(r)=\frac{\omega^2}{c^2}\left(1-\frac{N^2}{\omega^2}\right)\left(1-\frac{S_{l}^2}{\omega^2}\right),
\end{equation}
$S_{l}$ is the characteristic acoustic frequency
\begin{equation*}
S_{l}=\frac{l(l+1)c^2}{r^2},
\end{equation*}
$N$ is buoyancy frequency and also called as Brunt-V\"{a}is\"{a}l\"{a} frequency. It is expressed as
\begin{equation}\label{eq:N2}
   N^2\simeq\frac{g^2\rho}{p}(\nabla_{\rm ad}-\nabla+\nabla_{\mu}),
\end{equation}
where, $\nabla_{\rm ad}$, $\nabla$, and $\nabla_{\mu}$ are the adiabatic temperature gradient, the temperature gradient, and $\mu$-gradient, respectively.

For high-order p modes typically $\omega^2\gg N^2$, then $K$ can be approximately expressed as \citep[][Eq. (5.29) in page 80]{jcd2003}
\begin{equation}\label{eq:Krp}
K(r)\simeq\frac{1}{c^2}(\omega^2-S_{l}^2).
\end{equation}
The length of the wave vector $|\mathbf{k}|^2$ can be expressed as the sum of a radial component and a horizontal component, i.e.,
\begin{equation}\label{eq:Kkrkh}
|\mathbf{k}|^2\equiv k_{r}^2+k_{\rm h}^2.
\end{equation}
The radial component $k_r$
\begin{equation*}
k_{r}^2=K(r)=\frac{1}{c^2}(\omega^2-S_{l}^2).
\end{equation*}
The horizontal component $k_{\rm h}$  \citep[][Eq. (4.51) in page 64]{jcd2003}
\begin{equation}
k_{\rm h}^2=\frac{l(l+1)}{r^2}.
\end{equation}
Therefore, substitute $S_{l}$, $k_r$, and $k_{\rm h}$ into Equation \eqref{eq:Kkrkh} and obtain the dispersion relation of p-mode oscillations \citep[][Eq. (3.55) in page 52]{jcd2003}
\begin{equation}
\omega^2=c^2|\mathbf{k}|^2.
\end{equation}

According to the definition of phase velocity \citep[more description refer to][Eq. (15.16) of page 116]{Unno1989nosbook}:
\begin{equation}\label{eq:pv}
v_{\rm phase}\equiv\frac{\omega}{k},
\end{equation}
the propagation velocity of p-mode is $v_{\rm phase,p-mode}=c$, i.e., p-mode propagates with sound speed in stars.

The definition of acoustic depth \citep[][Eq. (3.228) in page 219]{Aerts2010}
\begin{equation}\label{eq:acoudp}
\tau(r)\equiv\int^R_{r}\frac{dr'}{c},
\end{equation}
presents the propagation time of oscillating wave (p-mode) from stellar surface to stellar inner position $r'=r$. Correspondingly, the propagation time from stellar surface to center is called as acoustic radius
\begin{equation}\label{eq:acour}
\tau_{0}=\int^R_{0}\frac{dr'}{c}.
\end{equation}
It represents the ``size" (i.e., acoustic size) of the p-mode propagation cavity.

For high-order g-mode typically $\omega^2\ll S^2_{l}$ and then $K$ can be approximately expressed as \citep[][Eq. (5.33) in page 81]{jcd2003}
\begin{equation}\label{eq:Krg}
K(r)\simeq\frac{1}{\omega^2}(N^2-\omega^2)\frac{l(l+1)}{r^2}.
\end{equation}
Similarly, for g-mode, 
the length of wave vector can be expressed as
\begin{equation*}
|\mathbf{k}|^2=k_{r}^2+k_{\rm h}^2=K(r)+k_{\rm h}^2=\frac{N^2}{\omega^2}\frac{l(l+1)}{r^2},
\end{equation*}
i.e.,
\begin{equation*}
\omega^2=\frac{N^2}{|\mathbf{k}|^2}\frac{l(l+1)}{r^2},
\end{equation*}
which is the dispersion relation of g-mode oscillations \citep[more detail description refer to][Eq. (33.18) in page 286]{Unno1989nosbook}. Correspondingly, the propagation velocity of g-mode ($v_{\rm phase,g-mode}$) is
\begin{equation}
v_{\rm phase,g-mode}=\frac{\omega^2}{\sqrt{l(l+1)}}\frac{r}{N}.
\end{equation}
It shows that the propagation velocity is directly proportional to the oscillation frequency $\omega^{2}$ and inversely proportional to the degree $\sqrt{l(l+1)}$.
Similar to the definition of acoustic depth $\tau(r)$ -- Equation \eqref{eq:acoudp}, the buoyancy depth can be defined as
\begin{equation}\label{eq:buodp}
\pounds(r)\equiv\int^R_{r}\frac{dr'}{v_{\rm phase,g-mode}}=\frac{\sqrt{l(l+1)}}{\omega^2}\int^R_{r}\frac{N}{r'}dr',
\end{equation}
which also represents the propagation time of oscillation wave, but for g-mode oscillations. Correspondingly, the buoyancy radius which presents the time of a g-mode oscillating wave propagating from stellar surface to center is defined as
\begin{equation}\label{eq:buor}
\pounds_0\equiv\frac{\sqrt{l(l+1)}}{\omega^2}\int^R_{0}\frac{N}{r'}dr'.
\end{equation}

Compared with acoustic depth (or radius; Equations \eqref{eq:acoudp} and \eqref{eq:acour}), the buoyancy depth (or radius; Equations \eqref{eq:buodp} and \eqref{eq:buor}) is not only related to the stellar structure ($N$ and $r$) but also related to the oscillation frequencies $\omega$ and their degrees $l$. Therefore, we adopt the quantities of
\begin{equation}\label{eq:buodpL}
\Lambda(r)=\int^R_{r}\frac{N}{r'}dr'
\end{equation}
and
\begin{equation}\label{eq:buorL}
\Lambda_0=\int^R_{0}\frac{N}{r'}dr'
\end{equation}
to replace $\pounds(r)$ and $\pounds_0$ to characterize the buoyancy size (buoyancy depth and radius) of stars. Similar to $\tau(r)$ and $\tau_{0}$,  $\Lambda(r)$ and $\Lambda_0$ are merely dependent on stellar structure and independence of the oscillations. Correspondingly, the dimensions of the improved buoyancy depth and radius are transformed to ones of angular frequency (radian per second) from ones of time.
Correspondingly, the buoyancy radius $\Lambda_0$ can be expressed with period spacing $\Delta\Pi_{l}$ as following
\begin{equation}\label{eq:buorL}
\Lambda_0=\frac{\pi}{\sqrt{l(l+1)}}\Delta\Pi^{-1}_{l}.
\end{equation}

\section{Appendix B: Inlist File of Pulse in MESA (V6208)}

\noindent \&star\_job  ! HD49385 \\

    \noindent create\_pre\_main\_sequence\_model = .true.\\
      kappa\_file\_prefix = 'gs98'\\
      change\_initial\_net = .true.\\
      new\_net\_name = 'o18\_and\_ne22.net'\\

\noindent / ! end of star\_job namelist\\

\noindent \&controls\\

\noindent      initial\_mass =      0.60875000D+01\\
      initial\_z =       0.43750000D-01\\
      initial\_y =       0.26625000D+00\\
      overshoot\_f\_above\_burn\_h =       0.20000000D-01\\
      min\_D\_mix =       0.50118723D+04\\

\noindent      do\_element\_diffusion = .false.  ! .true.\\

\noindent      calculate\_Brunt\_N2 = .true.\\
      !use\_brunt\_dlnRho\_form = .true.\\
      use\_brunt\_gradmuX\_form = .true. \\
	  which\_atm\_option = 'Eddington\_grey' \\

\noindent      max\_years\_for\_timestep = 0.1d6\\
      varcontrol\_target = 1d-4 ! for main sequence stars (5d-4 for pre\_main\_sequence)\\
      dH\_hard\_limit = 1d-3\\

\noindent      mesh\_delta\_coeff = 0.4\\
      max\_allowed\_nz =30000  ! maximum number of grid points allowed\\
      max\_model\_number = 70000 ! negative means no maximum\\
      xa\_central\_lower\_limit\_species(1) = 'h1'\\
      xa\_central\_lower\_limit(1) = 0.05\\
      mixing\_length\_alpha = 2. \\
      set\_min\_D\_mix =.true.\\
      min\_center\_Ye\_for\_min\_D\_mix = 0.4 ! min\_D\_mix is only used when center Ye $>$= this\\
      dH\_div\_H\_limit\_min\_H = 2d-1\\
      dH\_div\_H\_limit = 0.0005d0\\
      dH\_div\_H\_hard\_limit = 1d-2\\

\noindent      / ! end of controls namelist\\

\end{document}